\documentclass[conference]{IEEEtran}
\usepackage{cite}
\usepackage{amsmath,amssymb,amsfonts}
\usepackage{algorithmic}
\usepackage{graphicx}
\usepackage{hyperref}
\usepackage{textcomp}
\usepackage{xurl}
\usepackage{xcolor}
\usepackage{array}
\usepackage{tabu}
\usepackage{float}

\def\BibTeX{{\rm B\kern-.05em{\sc i\kern-.025em b}\kern-.08em
    T\kern-.1667em\lower.7ex\hbox{E}\kern-.125emX}}

\DeclareRobustCommand*{\IEEEauthorrefmark}[1]{%
  \raisebox{0pt}[0pt][0pt]{\textsuperscript{\footnotesize #1}}%
}

\begin{document}

\title{Quantum Picturalism:\\Learning Quantum Theory in High School}

\author{
  \IEEEauthorblockN{%
    Selma D\"undar-Coecke\IEEEauthorrefmark{1}$^,$\IEEEauthorrefmark{2},
    Lia Yeh\IEEEauthorrefmark{1}$^,$\IEEEauthorrefmark{3},
    Caterina Puca\IEEEauthorrefmark{1},
    Sieglinde M.-L. Pfaendler\IEEEauthorrefmark{4},\\
    Muhammad Hamza Waseem\IEEEauthorrefmark{1}$^,$\IEEEauthorrefmark{5},
    Thomas Cervoni\IEEEauthorrefmark{1},
    Aleks Kissinger\IEEEauthorrefmark{3},
    Stefano Gogioso\IEEEauthorrefmark{3}$^,$\IEEEauthorrefmark{6},
    Bob Coecke\IEEEauthorrefmark{1}
  }
  \IEEEauthorblockA{%
    \IEEEauthorrefmark{1}\textit{Quantinuum, 17 Beaumont Street, Oxford OX1 2NA, UK}\\ \texttt{\{bob.coecke $|$ selma.coecke $|$ thomas.cervoni $|$ caterina.puca\}@quantinuum.com} \\
    \IEEEauthorrefmark{2}\textit{Centre for Educational Neuroscience, UK}\\
    \IEEEauthorrefmark{3}\textit{Department of Computer Science, University of Oxford, UK}\\ \texttt{\{stefano.gogioso $|$ aleks.kissinger $|$ lia.yeh\}@cs.ox.ac.uk}\\
    \IEEEauthorrefmark{4}\textit{IBM Deutschland Research \& Development GmbH, Schönaicher Str. 220, D-71032 Böblingen, Germany}\\
    \IEEEauthorrefmark{5}\textit{Department of Physics, University of Oxford, UK}\\ \texttt{hamza.waseem@physics.ox.ac.uk}\\
    \IEEEauthorrefmark{6}\textit{Hashberg Ltd, London, UK}\\
  }
}%end author

\IEEEoverridecommandlockouts \IEEEpubid{\begin{minipage}{\textwidth}\ \\[12pt]\ \copyright2023 IEEE. Personal use of this material is permitted. Permission from IEEE must be obtained for all other uses, in any current or future media, including reprinting/republishing this material for advertising or promotional purposes, creating new collective works, for resale or redistribution to servers or lists, or reuse of any copyrighted component of this work in other works.\hfill \end{minipage}}

\maketitle

\begin{abstract}
Quantum theory is often regarded as challenging to learn and teach, with advanced mathematical prerequisites ranging from complex numbers and probability theory to matrix multiplication, vector space algebra and symbolic manipulation within the Hilbert space formalism. It is traditionally considered an advanced undergraduate or graduate-level subject.

In this work, we challenge the conventional view by proposing “Quantum Picturalism” as a new approach to teaching the fundamental concepts of quantum theory and computation. We establish the foundations and methodology for an ongoing educational experiment to investigate the question “From what age can students learn quantum theory if taught using a diagrammatic approach?”.
We anticipate that the primary benefit of leveraging such a diagrammatic approach, which is conceptually intuitive yet mathematically rigorous, will be eliminating some of the most daunting barriers to teaching and learning this subject while enabling young learners to reason proficiently about high-level problems.
We posit that transitioning from symbolic presentations to pictorial ones will increase the appeal of STEM education, attracting more diverse audience.
\end{abstract}


\section{Introduction}

Quantum theory is often regarded as one of the most inaccessible scientific disciplines.
Its introduction to university students is typically delayed to the later years of their training due to the complexity of the mathematical and physical content and a substantive reliance on symbolic thinking within the Hilbert space formalism~\cite{Marshman2016difficultqoperators}.
Meanwhile, the last few decades have seen multiple proposals to teach quantum science to students as young as secondary schoolers~\cite{stadermann2019analysis, michelini2000proposal, bitzenbauer2020new, escalada2004student}. However, despite the progress made by recent advancements~\cite{economou2022hello, walsh2021piloting, perry2019quantum, hughes2022teaching, davis2022quantum}, most current approaches to teaching quantum concepts at both high school and university levels remain hampered by a significant number of advanced mathematical prerequisites, including matrix multiplication, complex numbers, and probability theory. 
A recent survey of 28 introductory undergraduate courses in quantum information science, intended for physicists, computer scientists, and mathematicians, found that 75\% had linear algebra as a prerequisite~\cite{meyer2022interdisciplinary}.
Furthermore, instructors interviewed as part of the survey reported that their students had difficulty both at the mathematical level---linear algebra and complex numbers\footnote{Presently only part of the Further Maths syllabus in the UK.\newline \newline } being the main barriers to entry---and at the physical level.
% \textcolor{magenta}{Sieglinde note: ¿Do we have a copy of the latest syllabus? in 1998, there was an option to do further/advanced maths and standard maths. Only further maths covered complex numbers, so many had not seen complex numbers until they had joined university. What is the case now?  }
% \textcolor{magenta}{ The next paragraph about applying tools with conceptual understanding makes sense if we can have a way of distinguishing that they correctly applying the spider rules without conceptual understanding  vs with conceptual understanding. How would that be done?}
It has also been reported in the literature that students in quantum mechanics courses are ``often able to apply mathematical tools without a corresponding conceptual understanding''~\cite{baily2010teaching}---a learning outcome likely to have its roots in a historically widespread `shut up and calculate' culture within the quantum sciences~\cite{johansson2018shut,kaiser2014shut}.

The expectation for students to be proficient in such mathematical topics is carried into early efforts to teach quantum computing at the high school level~\cite{perry2019quantum}.
However, teaching strategies that purposefully avoid formal mathematical tools are reportedly associated with learning difficulties and misconceptions~\cite{krijtenburg2017insights}.
Hence, the educational challenge is to reform the teaching of quantum science by (1) reducing the complexity of the abstract mathematical formalism~\cite{bouchee2022towards}, while (2) enhancing the conceptual understanding and (3) maintaining a high level of scientific rigour.

The development of categorical quantum mechanics~\cite{abramsky2009categorical} in the early 2000s showed that an alternative approach to introductory quantum education was possible.
Diagrammatic methods were first introduced in 2005 at the University of Oxford, enabling quantum mechanical reasoning ``using only pictures of lines, squares, triangles and diamonds''~\cite{coecke2006kindergarten}, i.e. with visual, yet formal, tools.
% \textcolor{magenta}{ The sentence following this note is a little hard to read. Could it be rewritten into two shorter sentences? }
Such methods were introduced as the graphical counterpart to strongly compact closed categories, a kind of algebraic structure introduced shortly before~\cite{abramsky2004categorical}.
But the tongue-in-cheek title of the later work~\cite{coecke2006kindergarten}, ``Kindergarten Quantum Mechanics", clearly hinted at the possibility of future educational benefits for younger audiences.

The term ``Quantum Picturalism'' (QP) was coined in 2009~\cite{ContPhys}, when it was also suggested that---in the longer term and subject to further development---QP could make quantum physics accessible to a wider audience beyond university-level physics students. Specifically, it would remove the requirement to have a prerequisite in linear algebra and complex numbers, thus also making the concepts accessible to a younger high school audience. 
A first experiment in this regard was already sketched in~\cite[\S6]{ContPhys}, while a second experimental draft can be found in \cite{exp1}.

QP effects a paradigm shift in how quantum physics is understood and practiced using visually intuitive but mathematically rigorous diagrams, called string diagrams. These diagrams are interdisciplinary, having applicability not only in quantum physics but also in areas such as machine learning, control systems, electronics, game theory, linguistics, control theory etc. Therefore, these diagrams help forge unforeseen connections between disparate areas of STEM. One example is the emerging area of quantum natural language processing, which came into being by drawing an analogy between quantum physics and language.
Innovation occurs when subjects with apparently no connection talk to one another. This methodology (QP) shifts the focus towards a network mindset which fosters innovation.

The use of pictorial representations has a rich history in educational implementations, with extensive evidence thoroughly explored~\cite{tversky2004semantics}. There has been a major focus on the function of instructional pictures as valuable support for, or as an alternative to, both text and symbolic reasoning \cite{bobek2016creating, mielicki2015affordances, herrlinger2017pictures}
with evidence suggesting that visual aids like diagrams enhance learning by providing intuitive explanations, improving memory retention, and catering to diverse learning styles~ \cite{verdi1997organized}. A common way of using visual aids is to simplify complex concepts, while promoting creativity and problem-solving skills. Indeed, there is a growing interest in integrating visual aids into educational implementations, endorsed by modern pedagogical approaches, to foster enriched learning experience and enhance scientific thinking~\cite{mayer2005cambridge, carney2002pictorial, fan2015drawing}.
% \textcolor{cyan}{May need more refs here.} \textcolor{orange}{Better now?}
% \textcolor{cyan}{Does \cite{tversky2004semantics} belong here?}

Consistent with this notion, we claim that the entirely visual nature of QP enhances understanding of quantum theory by lowering the technical barrier, making the topic suitable for a high-school level introduction.
Contrary to visual aids such as multimedia and schematic diagrams (flowcharts, mindmaps, etc.), QP is a mathematically sound diagrammatic method, stemming from a well-established lineage of academic research on ``graphical calculi''~\cite{coecke2022kindergarden}. Furthermore, QP can be used as a self-contained language to describe quantum theory, eliminating the need for informal explanations or additional formal backing: it inherits all the educational benefits of visual approaches while also doubling as a descriptive and computational tool.
The rapidly increasing adoption of QP by both academia and the quantum industry further shows that such advantages come at no expense of scientific rigour or cutting-edge applicability.

\begin{figure*}[h!]
\centering
\includegraphics[width=0.69\textwidth]{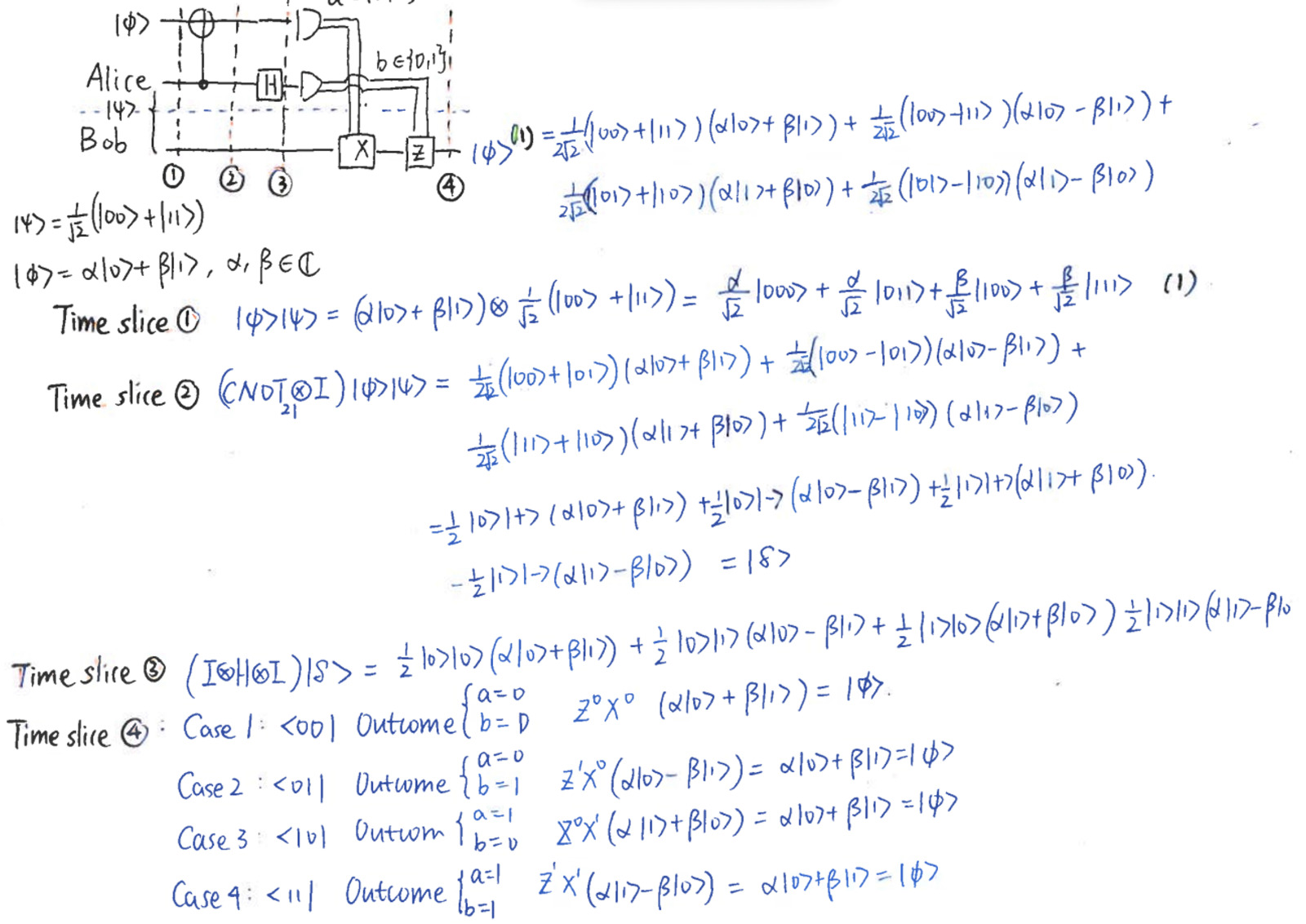}
\caption{
Calculations in bra-ket notation by Master's student Sarah Meng Li from the University of Waterloo, demonstrating the correctness of the quantum teleportation protocol as part of their graduate course QIC 710: Intro to Quantum Information Processing. The figure exemplifies how the Hilbert space formalism produces a much more arcane description of the quantum teleportation protocol, as opposed to the intuitive diagrammatic version of Fig. \ref{fig:diagrammatic-reasoning-teleportation-steps}.
}
\label{fig:teleportation-learning}
\end{figure*}

Contrary to the prevailing belief that symbolic mathematics is the language of physics, we posit that QP offers an approach to quantum theory that is superior for teaching and everyday practice.

Here, we aim to demonstrate how QP provides learners with effective tools for solving complex problems; and therefore it fosters engagement, improves self-confidence, builds motivation, and incorporates fun into the learning process. It is anticipated that such benefits will persist even in experiments where age, cognitive ability, science background, and attitude towards scientific topics are controlled for.

In response to the increased demand top-down through calls for workforce development, and bottom-up by students, recent years have seen growth of initiatives to teach quantum information science, quantum computing, and quantum mechanics to high school students~\cite{QxQ,OxPhys,IQCqsys,walsh2021piloting,economou2020teaching}.
Students can be attracted to quantum through games and interactive tools~\cite{seskir2022quantumgames, lacour2022vqol, Migdal2022qflytrap, chungyuan2022, Marshman2022, entanglementball2021}, programming~\cite{mykhailova2022, salehiseskir2022, qiskittextbook2021}, collaboration~\cite{khodaeifaal2022, maldonadoromo2022}, and art~\cite{quantumatlas,uchicago2022multidisciplinary}.
Likewise, various strategies have been developed to teach quantum theoretical concepts \cite{hoekzema2007particle, boe2023secondary, huseby2019observation, di2020development, rudolph2017, epiqc2020reversibility}.

There is an open debate as to which features best contribute to improving students' understanding of the subject matter~\cite{krijtenburg2017insights, greca2003does}.
However, the more formal approaches typically rely on symbolic mathematical foundations, using diagrams for descriptive and/or conceptual purposes rather than as the primary substrate for their delivery of knowledge.
In investigating the efficacy of QP as a teaching approach, we establish relevant baselines and gather experiential evidence for future studies in this area.

In this paper, we fully detail QP as a method to teach quantum theory to high school students, without prior exposure to the associated physics background nor the traditional mathematical prerequisites.

We introduce QP in Section~\ref{sec:QP}, and provide the details of a pilot experiment and evaluation methodology to test the efficacy of QP in education in Section~\ref{sec:xpmt}, followed by discussion in Section~\ref{sec:discuss}.
This adheres to the tradition in education sciences of preregistering experiments prior to their being conducted, to receive feedback on the methodology and to safeguard against irresponsible analysis of the data.

\IEEEpubidadjcol
\section{Quantum Picturalism}
\label{sec:QP}

Quantum Picturalism (QP) is an innovative approach to studying, teaching and practising quantum theory, relying on diagrammatic languages to conceptualise and manipulate highly abstract mathematical notions.

Its roots can be traced back to work by Penrose in the early 1970s, introducing early examples of graphical notation for tensors~\cite{Penrose}.
Work by Joyal and Street in the 1990s~\cite{JS} freed such graphical notations from the shackles of the physics that birthed them, introducing string diagrams as a language for a much broader realm of pure and applied mathematics within the unifying framework of category theory.
A specialisation of string diagrams to the algebraic structures of quantum theory in the early 2000s finally resulted in categorical quantum mechanics~\cite{abramsky2004categorical,coecke2006kindergarten,abramsky2009categorical}, the mathematical foundation for the QP approach.

QP is a multi-faceted framework with a variety of diagrammatic languages that capture quantum phenomena from different perspectives.
Of these languages, one of the oldest and most widely adopted is the ZX-calculus~\cite{coecke2011interacting}. Since its inception 12 years ago, the ZX-calculus has quickly found applications in quantum circuit optimisation~\cite{duncan2020graph, de2020fast, de2019techniques, kissinger2019reducing}, quantum error correction~\cite{huang2023qeczx, de2020zx, kissinger2022phase, khesinGraphicalQuantumCliffordencoder2023}, quantum simulation~\cite{kissinger2022classical}, quantum foundations~\cite{coecke2011phase, backens2016complete, gogioso2019dynamics, gogioso2017mermin}, measurement-based quantum computing~\cite{coecke2008interacting, duncan2009graph, kissinger2019universal}, quantum natural language processing~\cite{QPL-QNLP, lorenz2021qnlp, coecke2020foundations} and more. For learning resources on the ZX-calculus, adequate for all levels of familiarity with quantum theory, we refer to the ``Getting started'' section of \href{https://zxcalculus.com/#introPublications}{zxcalculus.com}.

Analogous diagrammatic approaches based on string diagrams have established themselves in several other scientific fields, including computer science~\cite{bonchi2014categorical}, linguistics~\cite{sadrzadeh2013frobenius, wang2023distilling}, and causal reasoning~\cite{lorenz2023causal}.
Several valuable variants of the ZX-calculus are used for quantum applications, such as the ZH-calculus~\cite{backens2018zh}, to study traditional quantum algorithms based on the quantisation of classical computation; and the ZXW-calculus~\cite{poor2023completeness}, with applications from photonic quantum computing~\cite{defelicelightmatterZXW} to Hamiltonian simulations and quantum chemistry~\cite{shaikh2022sum}.

The content and presentation for the course are based on the recently published book ``Quantum in Pictures''~\cite{coecke2022quantum}, by Coecke and Gogioso.
The book is an accessible version of the university-level textbook ``Picturing Quantum Processes: A First Course on Quantum Theory and Diagrammatic Reasoning''~\cite{CKbook}, by Coecke and Kissinger.
Material from ``Picturing Quantum Processes'' has been used for almost a decade at the University of Oxford for several courses in quantum theory offered by the Department of Computer Science.
In addition to being more accessible, engaging and reader-friendly, ``Quantum in Pictures'' is up-to-date with cutting-edge research, including content that did not even exist when ``Picturing Quantum Processes'' was published in 2017.

As an example to explain the differences between QP and traditional approaches to quantum theory, we now turn our attention to the quantum teleportation protocol.
Quantum teleportation is a peculiar effect---of interest in both quantum foundations and measurement-based quantum computing---where the state of a quantum system is transferred across arbitrary distances by encoding it into a string of classical bits.
The advantage of doing this lies in the observation that quantum systems are fragile and highly sensitive to noise, while classical bits can be stored, replicated, manipulated and transferred with ease.

\begin{figure*}[h]
\centering
\includegraphics[width=0.8\textwidth]{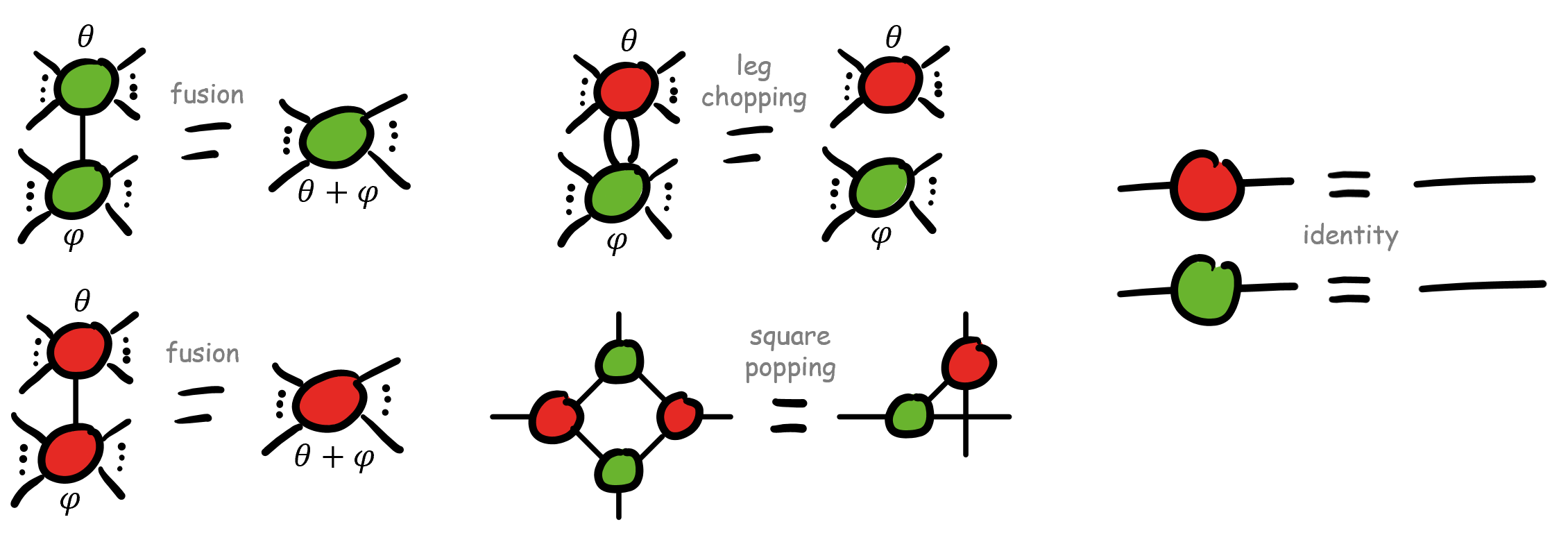}
\caption{A selection of diagrammatic substitution rules for the ZX calculus, the visual language used by the Quantum Picturalism approach to Quantum Theory.}
\label{fig:zx-rules}
\end{figure*}

Figure \ref{fig:teleportation-learning} above exemplifies how quantum teleportation is introduced in a typical quantum information course.
It showcases two salient features of the traditional Hilbert space approach: (1) the heavy reliance on symbolic algebra---typically, Dirac's bra-ket notation---for the explicit description of quantum systems and calculations associated with their transformations; and (2) the need to employ a separate graphical notation---typically, quantum circuit notation---to explain the spatiotemporal layout of said quantum system and keep track of the operations acting upon them.
The shortcoming of the often-used combination of bra-ket notation and quantum circuit notation, is that the quantum circuit notation shows the operations of the protocol whereas the bra-ket computation shows the final result, but there is missing a representation of the intermediate steps bridging the two notations. In contrast, the QP proof makes clear the role of the Bell state and Bell measurement: The colors indicate that the CNOT gate, the $|0\rangle$ state, and the $|+\rangle$ state exhibit interaction of the Z (green) and X (red) observables, a fundamental observation that is not made evident by the quantum circuit, bra-ket, or matrix formalism. This corroborates previous observation that students in physics ``have difficulties with representations of quantum operators corresponding to observables especially when using Dirac notation''~\cite{Marshman2016difficultqoperators}. In a sense, the QP formalism makes it immediately clear that various components of quantum circuits are made ``of the same stuff'', and therefore interact in interesting ways (hence the title of the first ZX-calculus paper, `Interacting Quantum Observables'~\cite{coecke2011interacting}).

Below is the QP description of the quantum teleportation protocol, written in the diagrammatic language of the ZX-calculus.
Only two diagrammatic ingredients are needed: red and green circles---known as ``spiders''---annotated by angles and connected by lines.
The angles---known as ``phases''---indicate the extent to which qubits are rotated by various parts of the protocol, while the lines---known as ``wires''---indicate the exchange of quantum information (physical or virtual).
\begin{center}
    \includegraphics[width=0.4\textwidth]{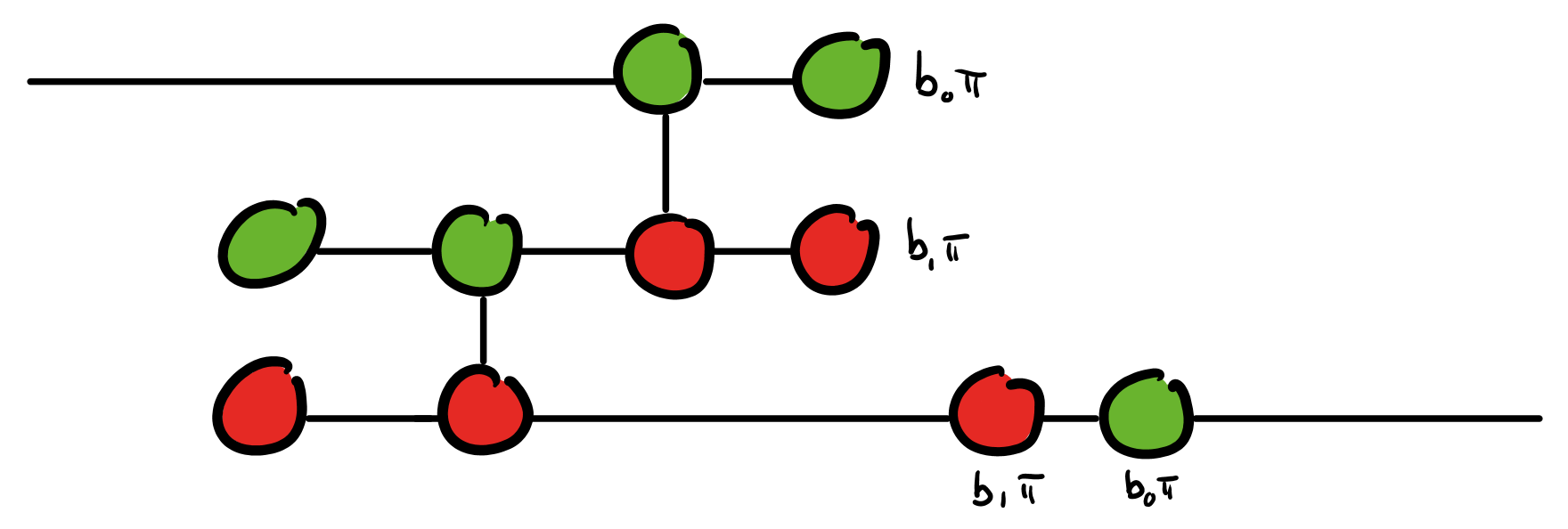}
\end{center}
Below is an annotated version of the same figure, indicating which parts of the ZX-diagram correspond to which parts of the traditional quantum circuit representation: the protocol takes a qubit as input (the wire on the left), results in a qubit as output (the wire on the right), and it involves two additional qubits, prepared in specified initial states (on the bottom left).
Two CNOT gates are applied to the three qubits, two of which are then measured (on the top right). A CNOT gate is a pair of a green spider and a red spider, connected by a vertical wire, with two wires on the left indicating the input qubits and two wires on the right indicating the output qubits.
The measurement outcomes, the bits $b_0$ and $b_1$, are then used to perform ``corrections'' on the remaining qubit in the form of two rotations (on the bottom right).
\begin{center}
    \includegraphics[width=0.45\textwidth]{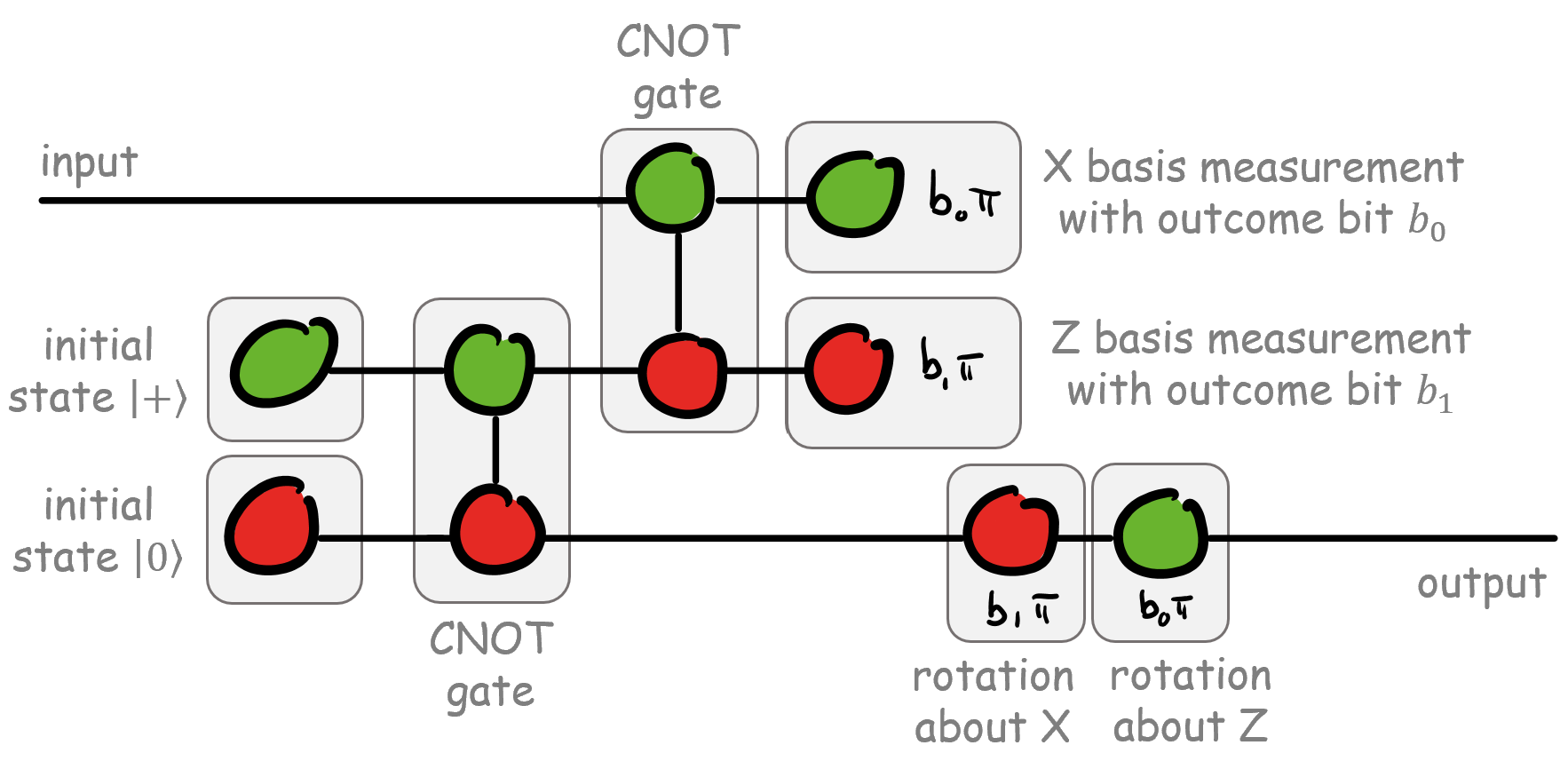}
\end{center}
Without knowing much about the ZX-calculus, a comparison with Fig. \ref{fig:teleportation-learning} already highlights one of the critical features of QP: It reveals that parts which appear very different in the quantum circuit representation---initial states, CNOT gates and measurements---are composed of the same basic building blocks---red and green spiders---wired together in different patterns.

\begin{figure*}
\centering
\includegraphics[width=\textwidth]{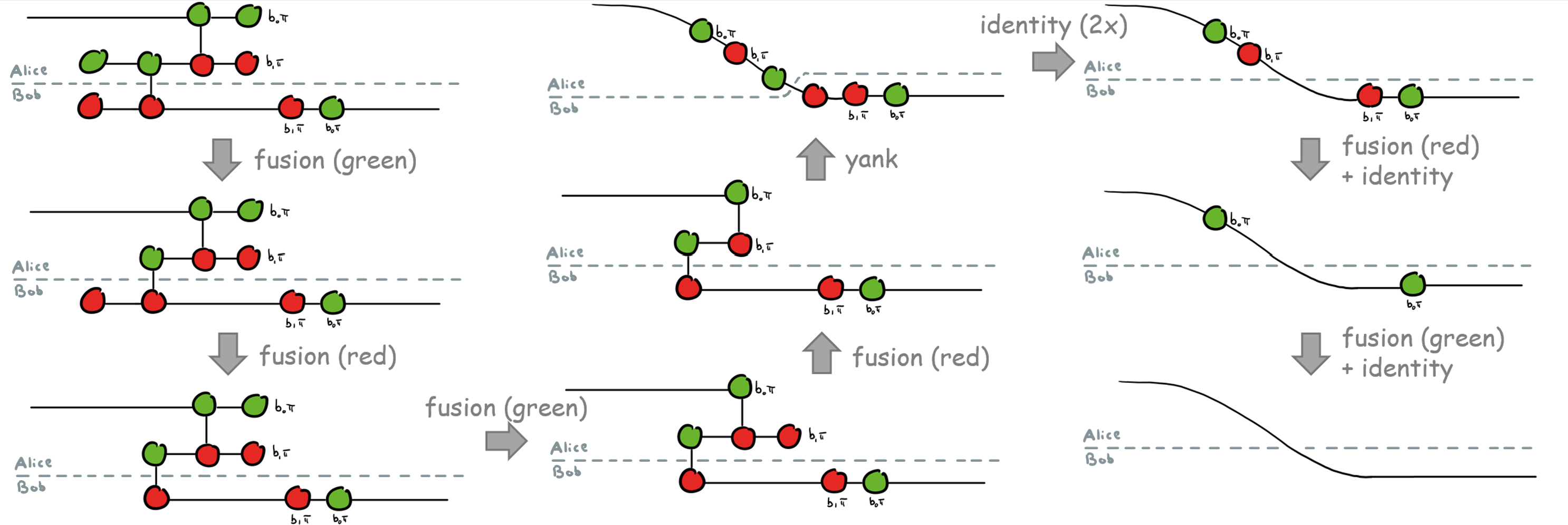}
\caption{
Proof of correctness for the quantum teleportation protocol. This is the QP analogue of the Hilbert space calculations from Fig. \ref{fig:teleportation-learning} (p.\pageref{fig:teleportation-learning}). Each step of the proof is labeled with the diagrammatic rule of the ZX-calculus in Fig.~\ref{fig:zx-rules} used.
This elucidates the flow of quantum information. Here, it is clearer each operation in the quantum teleportation protocol contributes to sending quantum information from Alice to Bob.
The most important point of conceptual understanding is that the corrections Bob must make are precisely those which nullify the errors randomly generated by Alice's measurement; in other words, Alice's and Bob's Z and X spiders fuse to become zero phase spiders (phases of $2\pi = 0$ are unlabeled by convention), which can be removed by the identity rule.
Furthermore, the step of yanking the wires straight highlights a property of information flow in spacetime, namely that information is encoded by how spiders are connected, rather than by their placement on paper.
}
\label{fig:diagrammatic-reasoning-teleportation-steps}
\end{figure*}

The intent of the quantum teleportation protocol can be stated by the following diagrammatic equation, where the protocol (the diagram on the left) is stated to have the same effect (the equal sign) as moving the quantum state, without change, from its input to its output (the diagram on the right).
\begin{center}
    \includegraphics[width=0.45\textwidth]{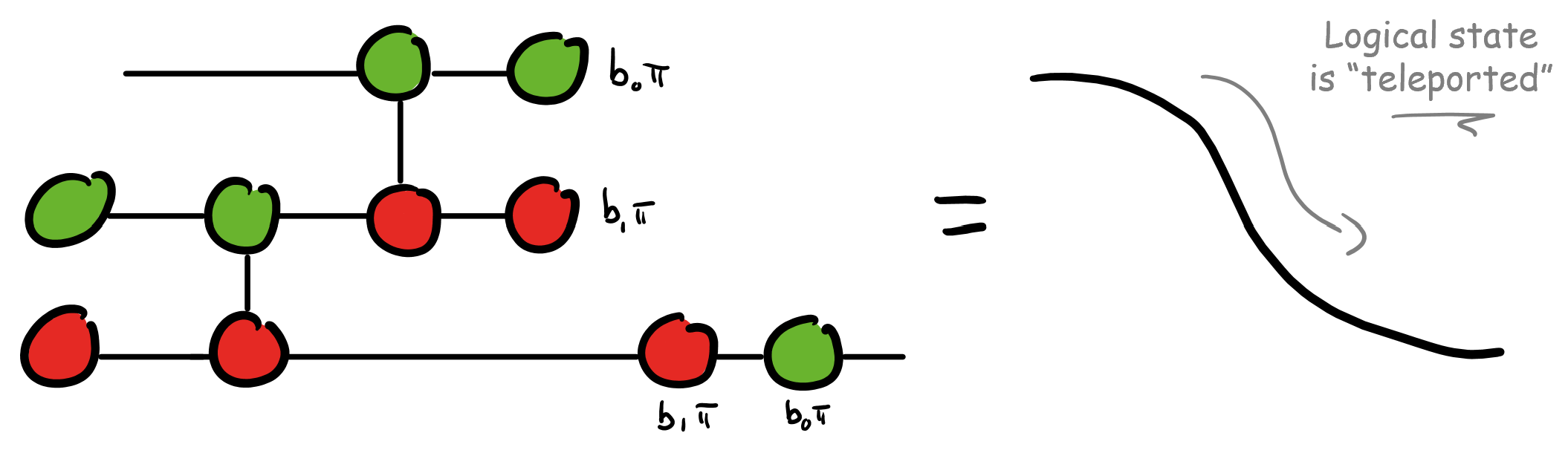}
\end{center}
In the QP approach, the same diagram that describes the protocol simultaneously provides the medium upon which any associated calculations can be performed.
This contrasts with the traditional Hilbert space approach, where different languages are used for calculation (bra-ket notation) and description (quantum circuit notation).
Calculations in the QP framework proceed by applying ``substitution rules'', graphical modifications of diagrams that maintain their meaning, i.e. that result in different diagrams with the same effect (symbolised by the equal sign).
Figure \ref{fig:zx-rules} (p.\pageref{fig:zx-rules}) presents some of the substitution rules for the ZX-calculus used by the examples in this paper. With the addition of only a few more other rules to the ones shown in Fig. \ref{fig:zx-rules}, the set of substitution rules is complete, i.e. suffices to make the ZX-calculus as expressive as the Hilbert space formalism for qubits~\cite{hadzihasanovic2018two} and more generally, for arbitrary finite dimension~\cite{poor2023completeness}.

As a warm-up exercise in diagrammatic substitution, below is the proof that two CNOT gates in sequence cancel each other out, leaving the two qubits---the two wires running left to right---unchanged.
\begin{center}
    \includegraphics[width=0.3\textwidth]{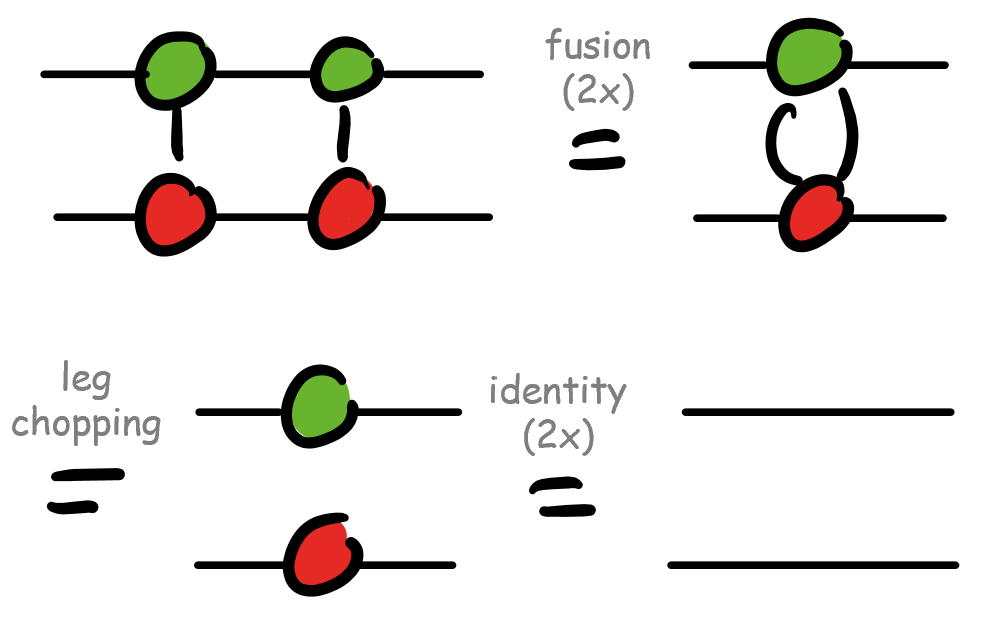}
\end{center}
The proof consists of two steps.
As the first step, we apply the ``fusion'' rule twice (cf. Fig. \ref{fig:zx-rules}, left): once to fuse the two green spiders, once to fuse the two red spiders.
As the second step, we apply the ``leg chopping'' rule (cf. Fig. \ref{fig:zx-rules}, middle top), to remove the pair of legs between the red and green spiders.
The results are two undecorated wires, carrying the two qubits from input to output unaltered.

% As a second warm-up exercise, below is the proof that three alternating CNOT gates have the same effect as swapping two qubits.
% This is an important practical observation in quantum computing, use to compile quantum circuits in architectures---such as superconducting quantum computers---where CNOT gates can only be performed between a small number of qubit pairs.
% \begin{center}
%     \includegraphics[width=0.25\textwidth]{Sections/pictures/triple-cx-three-lines.png}
% \end{center}
% The proof consists of two steps.
% As the first step, we apply the ``square popping'' rule (cf. Fig. \ref{fig:zx-rules}, middle bottom) to the rightmost two CNOTs, replacing them with a CNOT and a qubit swap.
% As the second step, we apply the equation from the previous exercise as a ``derived rule'', removing the two CNOTs and leaving the desired swap.

Now familiar with the diagrammatic substitution, we can focus on Figure \ref{fig:diagrammatic-reasoning-teleportation-steps} above, presenting the full proof of correctness for the quantum teleportation protocol.

The natively visual character of QP makes it easy to augment diagrams with other kinds of visual information, synergistically enhancing the communication power of all media involved.
For example, below is an explanation of how the fusion rule (cf. Fig. \ref{fig:zx-rules}, left) implements the action of rotation gates on quantum states.
\begin{center}
    \includegraphics[width=0.5\textwidth]{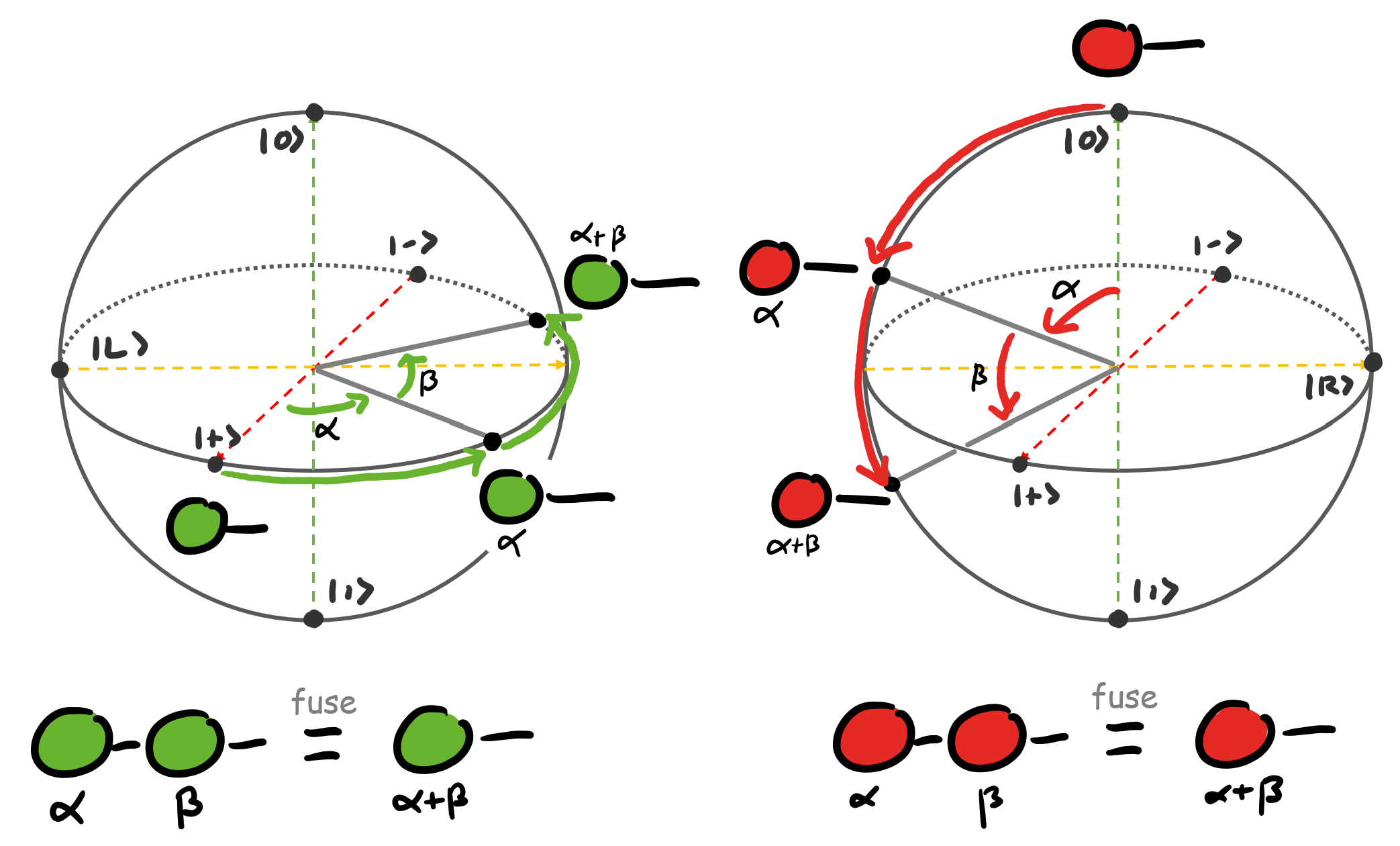}
\end{center}
The Bloch spheres endow the spiders with additional geometric meaning, locating them on the equator and prime meridian based on their phase.
The fusion rule, on the other hand, enhances the Bloch sphere presentation by adding a dynamical, computational description of the rotation action.

Diagrams are also easily augmented by non-technical conceptual art to convey meaning that exceeds the boundaries of mathematical and physical rigour.
For example, below is a diagrammatic representation of the Bell test, where an entangled state is measured simultaneously by two parties at a large distance apart.
\begin{center}
    \includegraphics[width=0.4\textwidth]{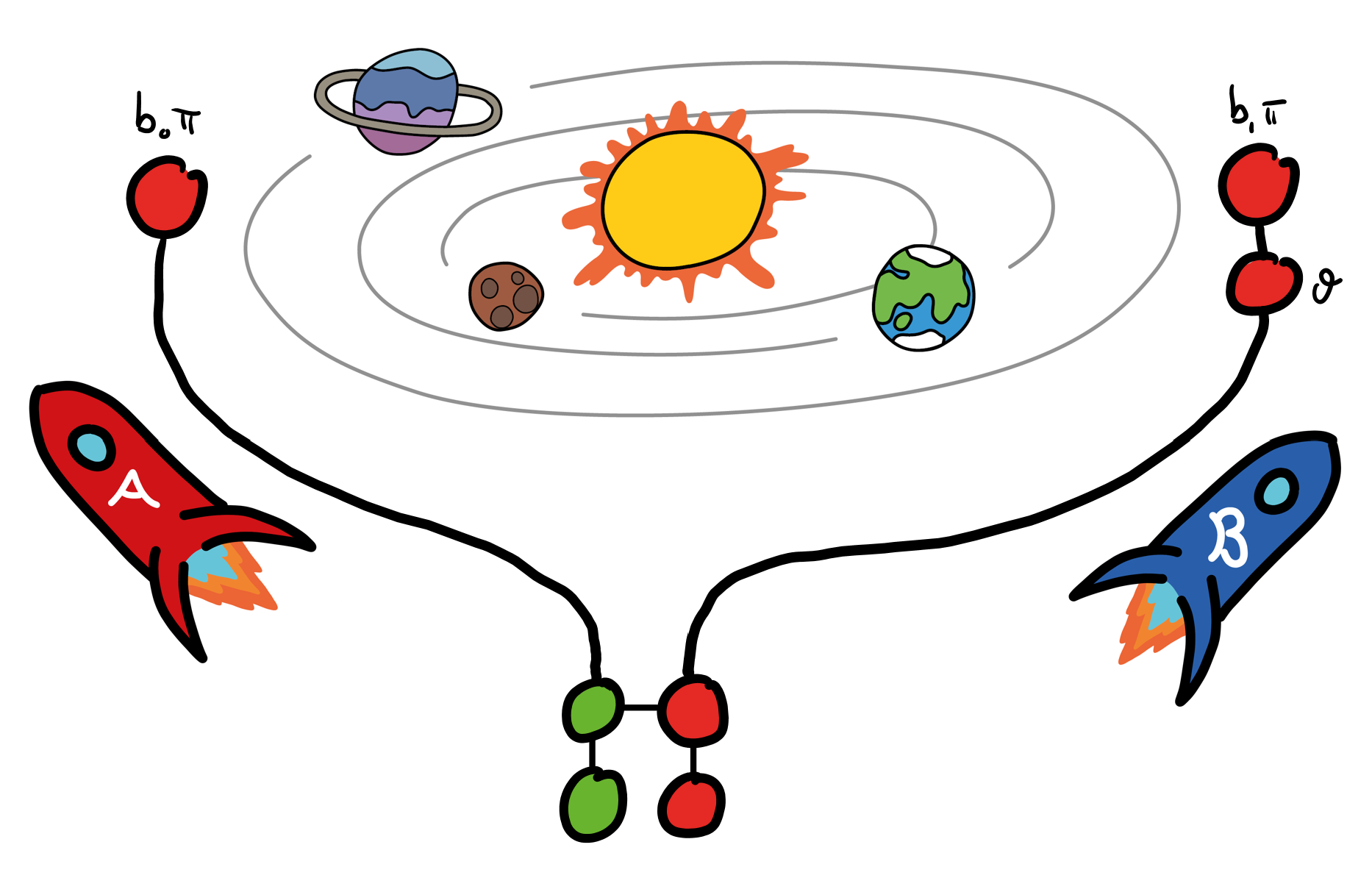}
\end{center}
The diagram is enriched by a number of artefacts and considerations.
Firstly, the diagram is drawn bottom-to-top\footnote{As opposed to the left-to-right convention, which we have previously adopted to represent the quantum teleportation protocol, matching the directional convention of quantum circuit notation.} to match worldline diagrams from general relativity.
Secondly, drawings of rockets and a solar system are used to indicate that, after becoming entangled, the two qubits are moved very far from each other.
Finally, the two measurements are horizontally aligned to indicate the (approximate) simultaneity of the operations performed by the two parties at a distance.

Finally, the diagrammatic representation itself can reveal hidden spatial structures within quantum algorithms, which may help explain how and why such algorithms work.
For example, below is a diagrammatic representation of the Quantum Approximate Optimisation Algorithm (QAOA), applied to solve the max-cut problem\footnote{An important combinatorial optimisation problem on networks. The max-cut problem is ``NP-complete'', i.e. solutions to this problem translate into solutions for a large class of problems of enormous real-world significance.} on a small network (6 nodes, corresponding to 6 qubits).
The traditional presentation of this algorithm (left below) uses initial states, CNOT gates, Z rotations and measurements, much as the quantum teleportation algorithm previously discussed.
By exploiting the visual flexibility of the QP approach, we repeatedly apply the fusion and square-popping rules (cf. Fig. \ref{fig:zx-rules}) and re-arrange the spider layout until the simplified diagram (right below) is obtained.
\begin{center}
    \includegraphics[width=0.48\textwidth]{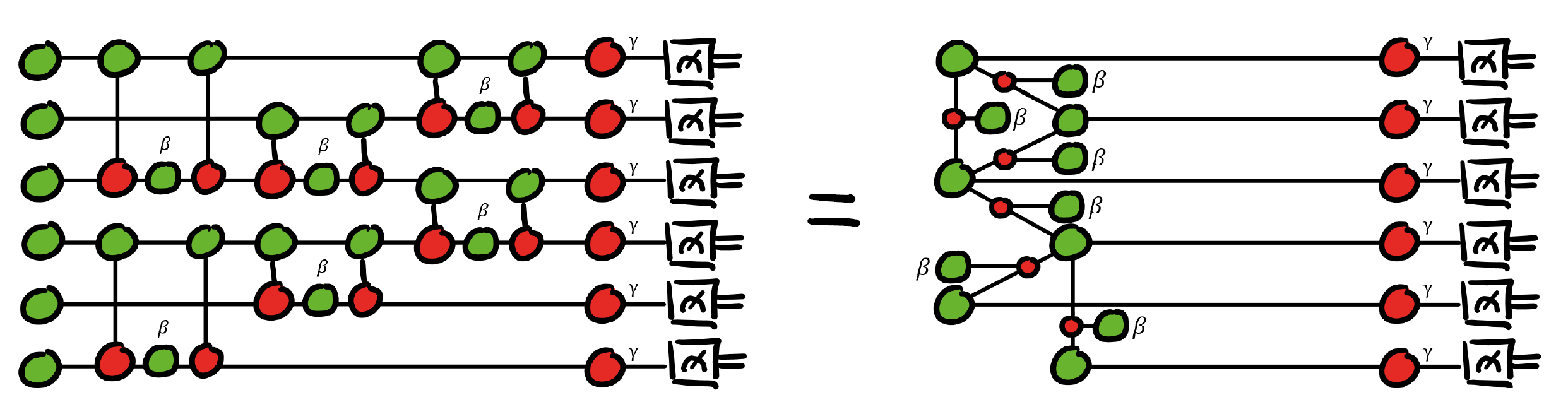}
\end{center}
The structure of the network (6 nodes connected by 6 edges, drawn in blue, left below) is revealed within the simplified and re-arranged diagram (spiders and wires highlighted blue, right below), exposing the mechanism that ultimately powers QAOA: the ability to entangle quantum systems in a way which matches the constraint pattern of the chosen optimisation problem.
The algorithm parameters---the $\beta$ and $\gamma$ angles below---allow the quality and degree of entanglement to be tuned, exploring the space of solutions in directions determined by the entanglement pattern.
\begin{center}
    \includegraphics[width=0.34\textwidth]{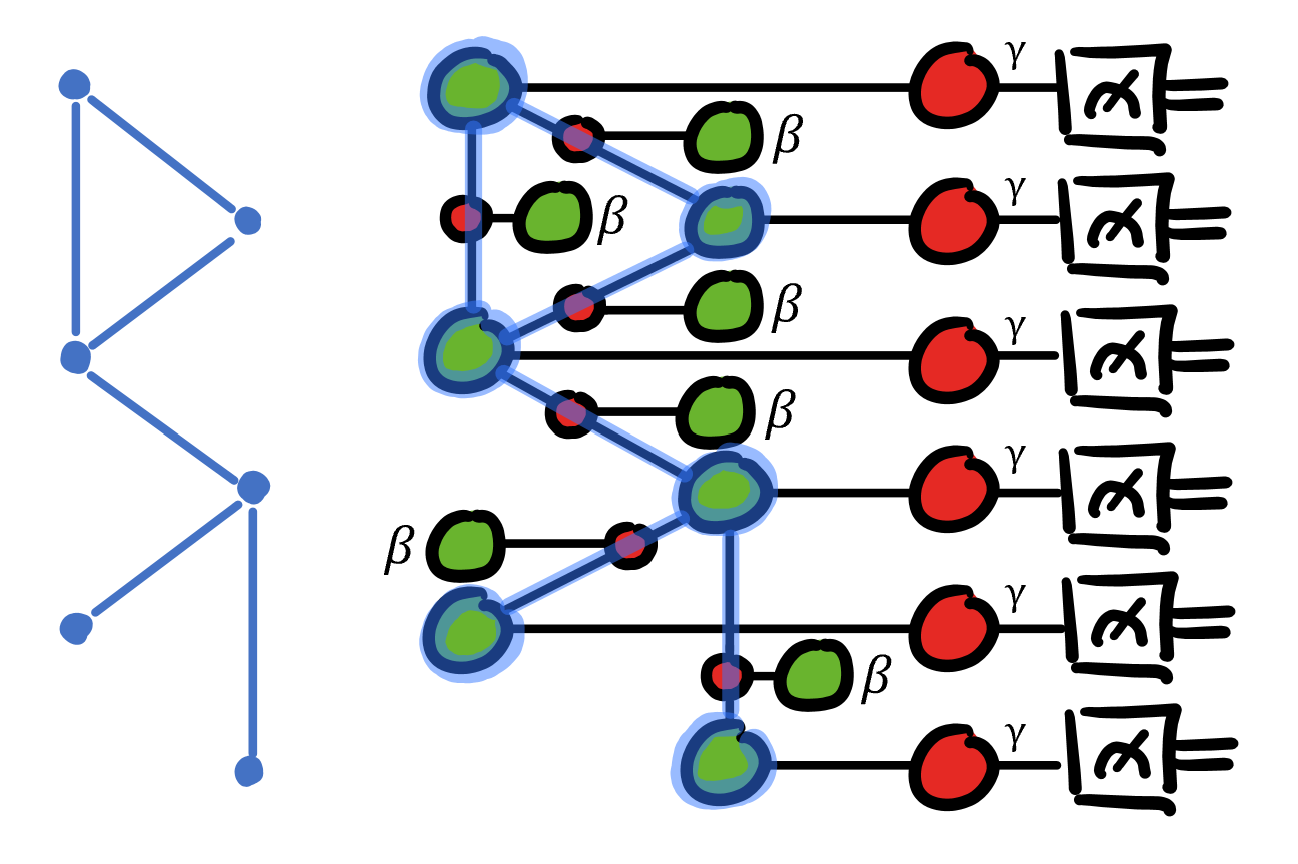}
\end{center}

% In analogy to the history of classical computer science, the Hilbert space formalism is to low-level assembly code what QP is to high-level programming languages. The Hilbert space formalism may have well been enough to get us through the early days of quantum computer science, but the exponential increase in complexity will soon necessitate higher-level methods, to bring forward applications that would be unimaginable with low-level conceptual tools alone.

\section{Experiment and evaluation}
\label{sec:xpmt}

This project was designed to collect evidence through an intervention trial to test whether (1) the diagrammatic formalism provides the most effective method to present quantum information theory (QIT) content, and (2) this formalism can significantly enhance young learners’ problem-solving ability after their training.

\subsection{Design and sampling}

This intervention trial is scheduled to be implemented between the 5th of June and the 11th of August, 2023.
It is designed to be delivered online through weekly one-hour sessions, each following a half an hour of self-study, spanning 10 weeks.
Through this trial, evidence has been 
 collected using a post-test design to evaluate the effectiveness of the QP approach.
A purposive sampling approach was employed to recruit 75 volunteer high-school students aged 16 to 19. The sample size was determined based on the number of available tutors. 

The project flyer was distributed to approximately 26 high schools across the UK and promoted on various social media platforms. Despite reaching out to a limited number of schools and teachers due to time constraints, we have received a more significant number of applications than anticipated $(N_0=734)$ within three weeks.
To ensure the quality of training, we  conducted an initial filtering process to exclude non-high-schoolers, respondents outside the target age range, and non-UK students as mandated by the obtained ethical approval.
We then asked for a study commitment of two hours weekly for 10 weeks, and sampled a random subset of positive respondents to reflect a balance of diversity criteria, re-sampling random subsets of respondents until all criteria were met within predetermined accuracy.
A total of 75 candidates were selected, with the goal of around 50 of them completing the course, taking into account potential attrition.

Participants have been provided with the experiment handbook that effectively communicates the objectives and requirements of the study; outlines the schedule; describes the evaluation methods to be used after the training; provides procedures for the tutorials; covers data protection and privacy policies; provides relevant contact details; and includes a handout which summarises the post-training procedures.
They have also been provided with a slimmed-down version of the QP book by Coecke and Gogioso~\cite{coecke2022quantum} to act as their tutorial book.
A similar and more substantive textbook by Coecke and Kissinger~\cite{CKbook} is actively  in use in various university-level courses, most notably at the University of Oxford.

All participants have been granted access, via a secure link, to weekly online recorded lectures (episodes) specifically designed to teach QP (a brief outline of the content is provided in Appendix A). They were also expected to engage in half-hour self-study sessions before the scheduled tutorials.
This was to help mitigate the effect of teacher and tutor variation on training outcomes, as each participant needed to have access to the same lectures during training via the project portal.

Participants have been provided with a link to book tutorials at their convenience. Each week's practice tutorials have been conducted by one of eight tutors, with a class size of 15 students maximum per tutorial.
 
To reduce the effect of tutor variation, the carousel technique was implemented: tutors rotated weekly, with a dedicated tutor being responsible for the week's material, enabling them to interact with each group at least once during the training period.
Consultation sessions with team members and tutors took place during the 1st, 3rd and 6th weeks to enable tutors to share their experiences and discuss the implementation of necessary measures.

Tutors underwent a Disclosure and Barring Service (DBS) check following UK regulations and completed a one-hour training session ahead of the trial. The training covered the study's objectives, effective interaction with students, conflict management, strategies for handling students with varying abilities, guidelines for taking observational notes for later qualitative data analyses, and techniques for encouraging students to progress towards the next steps.
It was made clear that the tutorials would primarily focus on practical exercises within interactive sessions.

Our tutors are highly qualified doctoral students, lecturers from the University of Oxford, and researchers from Quantinuum, all with substantial teaching experience.
Tutors were explicitly instructed to devise their own plans for implementing additional examples to reinforce the video lectures. These plans were supported by the Principal Investigators (PIs), who assisted in various aspects, including providing additional examples for each tutorial week. Their support ensured the curriculum remained up-to-date and incorporated research expertise and industry knowledge. 

\subsection{Data collection procedures}

Participants’ performance changes in learning outcomes (outlined in Appendix A) has been monitored via tutors' notes every week leading up to the post-tests. The weekly monitoring helped determine whether individuals exhibited observable improvement as a result of the training. For this, we utilised both directly related measures of performance and qualitative assessments of achievements. 
In addition to tutor notes, the exam and attitude questionnaires evaluated participants’ performance, providing comprehensive data to assess their progress and attitudes toward the training.  

\subsection {Assessment and evaluation}

The exam questionnaire used in this study is identical to the one previously implemented by the PIs in the quantum physics course for undergraduate students at the University of Oxford.
The questions were specified by the PIs for this project and subsequently reviewed by two project members to evaluate their applicability, appropriateness, language, presentation style, variability, and representativeness.

Questions were chosen to test the level of proficiency across target skills, ranging from calculation by diagrammatic manipulation to conceptual understanding of quantum information scenarios represented by the diagrams.

The selection process for each question involved careful consideration of various factors, including: 
\begin{itemize}
    \item objectives and aim of the study,
    \item desired outcomes,
    \item relevant literature,
    \item previous implementations with undergraduate students,
    \item the expertise of the PIs, and input from the research team, 
    \item a review process to assess the questions' clarity, relevance, and appropriateness for the study population. 
\end{itemize}

Upon completing the 8th week of training, participants were given a two-week period to complete and submit their answers to the exam questions.
The submitted work underwent a rigorous marking process managed by two experienced PIs, Coecke and Kissinger, who have been marking similar work for over a decade. They independently assessed students' work by following the provided criteria in Appendix B. This ensured fairness and accuracy, where each examiner independently assessed the submission without knowing the other's decision. In cases where a significant discrepancy of more than 9\% in the assigned grades occurs, another PI, Gogioso, was involved to review and facilitate further discussion until a consensus is reached.
This ensures consistency and transparency in the marking process, aims to provide participants with clear evaluation criteria and constructive feedback alongside their marks, and maintains consistency in the final grades.  

Participants' attitudes were assessed before the trial, focusing on three aspects: (1) their commitment, (2) interest in pursuing a STEM career, and (3) providing at least one specific reason for their interest in participating.
The post-intervention attitude assessment gathered feedback on the overall setup, satisfaction with the tutorials, any changes in their career plans, and their post-training/future expectations. Additionally, an optional question regarding their grades in STEM subjects was included to aid in analysing correlations between exam scores and attitude questions.
The attitude assessments generated qualitative data, which will be analyzed using qualitative analysis techniques in a follow-up study.

\subsection{Measured variables}

The outcome variables for the assessment, derived from the two questionnaires above (exam and attitude), focus on the number of problems students successfully solve using QP. 

The marking scheme prioritized comprehension rather than solely assessing the overall accuracy of answers. The evaluation process was guided by the following rubric, which is designed to align with the objectives, learning outcomes, and analysis scheme outlined in the breakdown marking criteria provided in Appendix A:

\begin{itemize}
\item Methodology: Assessing general problem-solving skills, including problem analysis, the ability to break down complex problems into smaller components, and logically organize the steps required to find a solution.
\item Correctness: Evaluating the accuracy of solutions and the validity of the reasoning provided in the answers.
\item Conciseness: Assessing the ability to present solutions concisely, considering factors such as the number of steps required to reach a solution.
\item Originality/Creativity: Considering the creativity and originality of the strategies employed by students in solving problems.
\item Evidence of understanding: Evaluating the correct application of relevant concepts and strategies taught during the course, such as the appropriate use of diagrammatic rewriting rules, and demonstrating a deep understanding of the physical meaning behind concepts like diagrammatic quantum teleportation.
\item Structuring, communication, presentation: Assessing the ability to organize, write, and draw solutions in a clear and unambiguous manner. This includes aspects such as the proper use of colors, distinguishing between ``quantum'' and ``ordinary'' diagrams, and using terminology consistent with the content presented in the tutorials (e.g., ``square-popping'', ``leg-chopping'', etc.).
\end{itemize}

Self-reports were also taken into account during the post-test period, in conjunction with the weekly observational data collected by tutors. Tutors received a feedback form they completed after each week's tutorial. This was to facilitate the assessment of each student's progress throughout the training and enable meaningful comparisons.

Further demographic factors, such as age, parental education, English language proficiency, and school type, were analysed as independent variables within the socio-economic and cultural impact on performance.

\subsection{Data exclusion rule}

Participants have been informed that they can withdraw from the study at any point.
Attendance during the training period was monitored, but only data from participants who attended tutorials and completed post-testing were included in the analysis.
As per the attendance rules agreed during the recruitment stage, participants had to register for one of five tutorial time options each week.
%Participants who were absent for more than one week of the online tutorials were excluded from the study.

To prevent the spillage  effect, participant interactions were discouraged during and after the training until they had submitted their exam answers. They were requested to keep their personal information private from each other to uphold the integrity of their training and ensure the validity of the analyses. With these precautions in place, no  unsupervised interaction or violation of research protocols between participants was observed  that would require us to exclude their data from the analyses, ensuring the study's validity. 

\subsection{Data Transformations}

For privacy reasons, raw subject data stored in a single/secure location accessible only to the project PI, Kissinger.
Students were then assigned a unique code to enable anonymised data to be shared and analysed independently by research team members. 
The researchers worked together throughout the data collection process, following the protocols for test administration. 
As a follow-up, frequency tables generated for each variable using SPSS / R to check that there are no out-of-range values and that any missing data points have been appropriately coded.
This strategy of creating single intermediate records, collating all raw data relating to each participant, has proved to be an essential facilitator of data checking and cleaning before entry for analysis, and tabulated data can be set up to allow for semi-automated entry once this process has been completed. 
Once data collection was ongoing, the researchers double-checked the processes involved, including coding and storing the raw data from the testing.
The coding schemes were modified where necessary.

\subsection{Analysis}

The impact of training was measured via post-tests.
Effects were assessed using analysis of variance. Differences in performance between the age groups were assessed using hierarchical regressions, path analyses, and multilevel modelling, also aiming to examine the influence of socioeconomic and demographic factors. 
Furthermore, underlying attitudinal factors examined using exploratory factor analysis.

\section{Discussion}
\label{sec:discuss}

Recently, there has been a resurgence of interest in exploring the quantum foundations of the universe, accompanied by a notable acceleration in the advancement of practical quantum technologies---a phenomenon sometimes known as the ``second quantum revolution''~\cite{dowling2003quantum}. However, a sound understanding of quantum theory has yet to penetrate mainstream awareness, education, and the professional training pipeline.

To make quantum knowledge accessible to a broader audience, we must  present complex scientific phenomena in a way that is more readily suitable for learning. This requirement necessitates a deviation from traditional approaches, which rely too heavily on complex symbolic reasoning and highly technical language. Additionally, our presentation should be practically applicable, as opposed to approaches which rely on inaccurate metaphors to convey a simplified, intuitive understanding of a complex subject matter. The experiment detailed in this paper proposes QP \cite{ContPhys} as a viable candidate for such a presentation, in the form of a visual paradigm already adopted in the wild for lecturing \cite{CKbook, coecke2023basic}, reasoning \cite{van2020zx}, and research \cite{de2020fast, de2019techniques, kissinger2019reducing, de2020zx, kissinger2022phase, khesinGraphicalQuantumCliffordencoder2023}.

This experiment is unique in three aspects.
Firstly, it offers state-of-the-art educational material, tailored to empower participants to perform quantum calculations using the diagrammatic formalism. Rather than merely advocating a `shut-up-and-calculate' approach, the curriculum presents a new conceptual foundation for learning physics and science, both quantum and beyond, with primary focus on a holistic understanding of the relationship between events and processes.
This unique perspective not only simplifies the formalism, but also supercharges the users' reasoning skills about quantum systems and their otherwise puzzling behaviours.

Secondly, the experiment aims to assess whether QP can democratise access to a crucial segment of university-level education by extending its reach to high school students. For almost a decade, QP has been taught to a small selection of students at elite universities, most notably the University of Oxford. As part of our experiment, a more diverse group of students---in terms of academic background, gender, socioeconomic status, and prior mathematical knowledge---will reap the benefit of QP for the first time.

Thirdly, the experiment explores whether learning quantum theory falls within the zone of proximal development of high school students, i.e. whether adequate guidance and support can enhance their understanding of such a subject matter.
A successful experiment would settle the question in the positive, while an unsuccessful outcome leaves the possibility that the experimental setup did not provide sufficient guidance to fully bridge the gap between the high-school syllabus and the more specialised requirements of quantum disciplines.

More evidence will be needed to test against diverse control groups, and to explore the effects of QP on larger samples of the high-school student population.
The facilitating effect of QP on the acquisition of quantum technology skills will also require further evaluation, investigating how it assists novices in accurately organising their mental representations while minimising the cognitive resources needed to achieve proficiency in real-world tasks.

The insights gained from this and future work will undoubtedly pave the way for the development of more inclusive and innovative educational endeavors. These efforts will play a significant role in bridging the gap between high school and university level learning, particularly for those aspiring to pursue a STEM career. It will help them establish crucial conceptual and abstract frameworks earlier, significantly contributing to their academic growth and understanding. 

Finally, the findings have the potential to inspire further research and advancement in quantum science pedagogy, contributing to the ongoing evolution of educational methodologies and practices specific to quantum technology and beyond.

\appendices

\section{Content and learning outcomes}

Below is the detailed week-by-week breakdown of course content and learning outcomes:

\begin{itemize}
\item In week 1, the episode ``Quantum in Pictures: Wires and Boxes'' introduces diagrams composed of wires and boxes to describe processes and perform mathematical operations. The episode also introduces the concepts of space and time in diagrams and the physics they represent.
\item In week 2, the episode ``Quantum Teleportation'' introduces the concept of the quantum lottery and the idea that wires and boxes can be used to carry out advanced mathematical reasoning. It then discusses quantum teleportation, a fundamental building block of quantum communication and quantum computing.
\item In week 3, the episode ``A World of Spiders'' introduces ``spiders'', a special kind of box with basic rules to work with them. Spiders---discussed in Section II on Quantum Picturalism---provide the fundamental building blocks for quantum computing.
\item In week 4, the episode ``Quantum Computing'' introduces quantum computing using spiders: logical operations (such as copying and adding), quantum gates (such as rotations, CNOT gates and Hadamard gates) and measurements.
It covers the use of diagrammatic substitution rules for circuit simplification and other relevant calculations.
% Classical computers use bits, and quantum computers use qubits. The episode covers the representation of bits and qubits using spiders "decorated" with phases. The episode also covers logical operations on those (qu)bits, like copying, adding, and the CNOT gate using spiders. For example, the CNOT gate is composed of a green spider and a red spider connected to each other---discussed in Section II on Quantum Picturalism. Participants also learn how the Hadamard gates (colour-changing boxes), introduced in the previous episode, affect a circuit of spiders. Learning how applying rules on spiders helps realise computations (\emph{i.e.} given quantum states at the beginning of a circuit, what are the quantum states at the end of the circuit, and the output of the computation) and simplify circuits, valuable to experimentalists with limited resources.
\item In week 5, the episode ``Quantum Teleportation with Spiders'' explores quantum teleportation further. It then introduces measurement-based quantum computing (MBQC), a fault-tolerant flavour of quantum computing which uses ideas from quantum teleportation to shift computation from gates to measurements. MBQC was also one of the motivating examples in developing the ZX-calculus.
\item In week 6, the episode ``Keeping Einstein Happy'' introduces notions of relativistic causality within the diagrammatic framework in the form of Sure-boxes and Maybe-boxes. 
\item In week 7, the episode ``Quantum vs Ordinary Particles'' introduces the concept of double wires to distinguish the quantum world from the classical world. Mathematically, doubling the wire gives new kinds of numbers (real probabilities, as opposed to complex amplitudes).
\item In week 8, the episode ``Everything Just in Pictures'' introduces the square-popping rule and the necessary tools to describe all quantum processes using pictures alone.
\item In week 9, participants receive a take-home exam, which they are requested to work on until the end of week 10.
\end{itemize}

\section{Marking Criteria}

The marking criteria are set out in Table \ref{tab:marking_criteria} (p.\pageref{tab:marking_criteria}).
They are designed to focus on specific assessment areas, ensuring alignment with the learning objectives and expected outcomes.
They are used in conjunction with discipline-specific criteria, and they should be viewed as guidance on the overall standards expected at different grade bands, aligning with the taught postgraduate generic marking criteria used at the University of Oxford.
The marking criteria have been reviewed and utilise a 0-100\% grading structure in line with the current university regulations.

\renewcommand{\arraystretch}{1.5}
\makeatletter
\renewcommand{\fnum@figure}{Table \thefigure}
\makeatother

\begin{figure*}[!t]
    \centering
\tabulinesep=1mm
\begin{tabu}{|X[c]|X[c]|X[c]|X[1.5c]|}
\hline
\multicolumn4{|c|}{\textbf{Distinction $>$70}} \\
\hline
\textit{Understanding} & \textit{Use of knowledge} & \textit{Structure} & \textit{Grade bands} \\
\hline
Advanced, in-depth, authoritative, full understanding of key ideas. Originality of the solutions, legitimacy of chain of reasoning in the answers provided.
&
Complex work and key problems solved. Correct application of concepts and techniques (e.g. the appropriate use of diagrammatic rewriting rules), the ability to use proper terminology (e.g. ``square-popping'', ``leg-chopping'')
&
Coherent and compelling work.
Logical and concise presentation. The solution drawn/written in a clear and unambiguous way (e.g. the proper use of notations, the difference between ``quantum'' and ``ordinary'' diagrams).
&
\textbf{(90-100)} insightful work displaying in-depth knowledge. Outstanding work, independent thought, highest standards of problem solving

\medskip

\textbf{(80-89)} insightful work displaying in-depth knowledge. Good quality of work, independent thought 

\medskip

\textbf{(70-79)} thoughtful work displaying in-depth knowledge, good standards of problem solving 
\\
\hline
\multicolumn4{|c|}{\textbf{Merit 60-69}} \\
\hline
\textit{Understanding} & \textit{Use of knowledge} & \textit{Structure} & \textit{Grade bands} \\
\hline
In-depth understanding of key ideas with evidence of some originality
&
Key problems solved. Correct application of most concepts, techniques, and correct use of terminology
&
Coherent work, logically presented. Clear solutions
&
\textbf{(65-69)} thoughtful work displaying good knowledge and accuracy. Evidence for the ability to solve problems

\medskip

\textbf{(60-64)} work displays good knowledge, some evidence for problem solving
\\
\hline
\multicolumn4{|c|}{\textbf{Pass 50-59}} \\
\hline
\textit{Understanding} & \textit{Use of knowledge} & \textit{Structure} & \textit{Grade bands} \\
\hline
Understanding of some key ideas with evidence of ability to reflect critically 
&
Some key problems solved. Correct application of some concepts, techniques, and terminology
&
Competent work in places but lacks coherence 
&
\textbf{(55-59)} work displays some understanding in most areas, but standard of work is variable

\medskip

\textbf{(50-54)} work displays knowledge and understanding of some areas, but some key problems are not solved
\\
\hline
\multicolumn4{|c|}{\textbf{Fail 40-49}} \\
\hline
\textit{Understanding} & \textit{Use of knowledge} & \textit{Structure} & \textit{Grade bands} \\
\hline
Superficial understanding of some key ideas, lack of focus 
&
Key problems are not solved/understood, gaps in application of concepts, techniques, and terminology
&
Weaknesses in structure and/or coherence 
&
\textbf{(40-49)} work displays patchy knowledge and understanding, most key problems are not solved
\\
\hline
\multicolumn4{|c|}{\textbf{Fail 0-39}} \\
\hline
\textit{Understanding} & \textit{Use of knowledge} & \textit{Structure} & \textit{Grade bands} \\
\hline
Lack of understanding
&
Key problems misunderstood/unanswered, limited/incorrect application of concepts, techniques and terminology 
&
Work is confused and incoherent 
&
\textbf{(33-39)} incomplete answers with some superficial knowledge

\medskip

\textbf{(20-32)} some attempt to write something relevant but many flaws

\medskip

\textbf{(0-19)} serious errors, irrelevant answers
\\
\hline
\end{tabu}
    \caption{Marking Criteria}
    \label{tab:marking_criteria}
\end{figure*}

\clearpage
\bibliographystyle{IEEEtran}
\bibliography{IEEEabrv,bibliography}

% Generated by IEEEtran.bst, version: 1.14 (2015/08/26)
\begin{thebibliography}{10}
\providecommand{\url}[1]{#1}
\csname url@samestyle\endcsname
\providecommand{\newblock}{\relax}
\providecommand{\bibinfo}[2]{#2}
\providecommand{\BIBentrySTDinterwordspacing}{\spaceskip=0pt\relax}
\providecommand{\BIBentryALTinterwordstretchfactor}{4}
\providecommand{\BIBentryALTinterwordspacing}{\spaceskip=\fontdimen2\font plus
\BIBentryALTinterwordstretchfactor\fontdimen3\font minus
  \fontdimen4\font\relax}
\providecommand{\BIBforeignlanguage}[2]{{%
\expandafter\ifx\csname l@#1\endcsname\relax
\typeout{** WARNING: IEEEtran.bst: No hyphenation pattern has been}%
\typeout{** loaded for the language `#1'. Using the pattern for}%
\typeout{** the default language instead.}%
\else
\language=\csname l@#1\endcsname
\fi
#2}}
\providecommand{\BIBdecl}{\relax}
\BIBdecl

\bibitem{Marshman2016difficultqoperators}
E.~Marshman and C.~Singh, ``Student difficulties with representations of
  quantum operators corresponding to observables,'' in \emph{2016 Physics
  Education Research Conference Proceedings}.\hskip 1em plus 0.5em minus
  0.4em\relax American Association of Physics Teachers, dec 2016.

\bibitem{stadermann2019analysis}
H.~Stadermann, E.~van~den Berg, and M.~Goedhart, ``Analysis of secondary school
  quantum physics curricula of 15 different countries: Different perspectives
  on a challenging topic,'' \emph{Physical Review Physics Education Research},
  vol.~15, no.~1, p. 010130, 2019.

\bibitem{michelini2000proposal}
M.~Michelini, R.~Ragazzon, L.~Santi, and A.~Stefanel, ``Proposal for quantum
  physics in secondary school,'' \emph{Physics Education}, vol.~35, no.~6, p.
  406, 2000.

\bibitem{bitzenbauer2020new}
P.~Bitzenbauer and J.-P. Meyn, ``A new teaching concept on quantum physics in
  secondary schools,'' \emph{Physics Education}, vol.~55, no.~5, p. 055031,
  2020.

\bibitem{escalada2004student}
L.~T. Escalada, N.~S. Rebello, and D.~A. Zollman, ``Student explorations of
  quantum effects in {LED}s and luminescent devices,'' \emph{The Physics
  Teacher}, vol.~42, no.~3, pp. 173--179, 2004.

\bibitem{economou2022hello}
S.~E. Economou and E.~Barnes, ``Hello quantum world! a rigorous but accessible
  first-year university course in quantum information science,'' \emph{arXiv
  preprint arXiv:2210.02868}, 2022.

\bibitem{walsh2021piloting}
J.~A. Walsh, M.~Fenech, D.~L. Tucker, C.~Riegle-Crumb, and B.~R. La~Cour,
  ``Piloting a full-year, optics-based high school course on quantum
  computing,'' \emph{Physics Education}, vol.~57, no.~2, p. 025010, 2021.

\bibitem{perry2019quantum}
A.~Perry, R.~Sun, C.~Hughes, J.~Isaacson, and J.~Turner, ``Quantum computing as
  a high school module,'' \emph{arXiv preprint arXiv:1905.00282}, 2019.

\bibitem{hughes2022teaching}
C.~Hughes, J.~Isaacson, J.~Turner, A.~Perry, and R.~Sun, ``Teaching quantum
  computing to high school students,'' \emph{The Physics Teacher}, vol.~60,
  no.~3, pp. 187--189, 2022.

\bibitem{davis2022quantum}
N.~A. Davis and B.~R. La~Cour, ``Quantum computing for the faint of heart,'' in
  \emph{2022 IEEE International Conference on Quantum Computing and Engineering
  (QCE)}.\hskip 1em plus 0.5em minus 0.4em\relax IEEE, 2022, pp. 669--672.

\bibitem{meyer2022interdisciplinary}
J.~C. Meyer, G.~Passante, S.~J. Pollock, and B.~R. Wilcox, ``The
  interdisciplinary quantum information classroom: Themes from a survey of
  quantum information science instructors,'' \emph{arXiv preprint
  arXiv:2202.05944}, 2022.

\bibitem{baily2010teaching}
C.~Baily and N.~D. Finkelstein, ``Teaching and understanding of quantum
  interpretations in modern physics courses,'' \emph{Physical Review Special
  Topics-Physics Education Research}, vol.~6, no.~1, p. 010101, 2010.

\bibitem{johansson2018shut}
A.~Johansson, S.~Andersson, M.~Salminen-Karlsson, and M.~Elmgren, ``“shut up
  and calculate”: The available discursive positions in quantum physics
  courses,'' \emph{Cultural Studies of Science Education}, vol.~13, pp.
  205--226, 2018.

\bibitem{kaiser2014shut}
D.~Kaiser, ``{History: Shut up and calculate!}'' \emph{Nature}, vol. 505, pp.
  153--155, 2014.

\bibitem{krijtenburg2017insights}
K.~Krijtenburg-Lewerissa, H.~J. Pol, A.~Brinkman, and W.~Van~Joolingen,
  ``Insights into teaching quantum mechanics in secondary and lower
  undergraduate education,'' \emph{Physical review physics education research},
  vol.~13, no.~1, p. 010109, 2017.

\bibitem{bouchee2022towards}
T.~Bouch{\'e}e, L.~de~Putter-Smits, M.~Thurlings, and B.~Pepin, ``Towards a
  better understanding of conceptual difficulties in introductory quantum
  physics courses,'' \emph{Studies in Science Education}, vol.~58, no.~2, pp.
  183--202, 2022.

\bibitem{abramsky2009categorical}
S.~Abramsky and B.~Coecke, ``Categorical quantum mechanics,'' \emph{Handbook of
  quantum logic and quantum structures}, vol.~2, pp. 261--325, 2009.

\bibitem{coecke2006kindergarten}
B.~Coecke, ``Kindergarten quantum mechanics: Lecture notes,'' in \emph{AIP
  Conference Proceedings}, vol. 810, no.~1.\hskip 1em plus 0.5em minus
  0.4em\relax American Institute of Physics, 2006, pp. 81--98.

\bibitem{abramsky2004categorical}
S.~Abramsky and B.~Coecke, ``A categorical semantics of quantum protocols,'' in
  \emph{Proceedings of the 19th Annual IEEE Symposium on Logic in Computer
  Science, 2004.}\hskip 1em plus 0.5em minus 0.4em\relax IEEE, 2004, pp.
  415--425.

\bibitem{ContPhys}
B.~Coecke, ``Quantum picturalism,'' \emph{Contemporary Physics}, vol.~51, pp.
  59--83, 2009, {a}rXiv:0908.1787.

\bibitem{exp1}
B.~Coecke, S.~D\"undar-Coecke, and S.~Gogioso, ``{IMPACT} kindergarten quantum
  theory,'' 2017,
  https://www.cs.ox.ac.uk/people/bob.coecke/Diag-QT-Exp.-Design.pdf.

\bibitem{tversky2004semantics}
B.~Tversky, ``Semantics, syntax, and pragmatics of graphics,'' 2004.

\bibitem{bobek2016creating}
E.~Bobek and B.~Tversky, ``Creating visual explanations improves learning,''
  \emph{Cognitive research: principles and implications}, vol.~1, pp. 1--14,
  2016.

\bibitem{mielicki2015affordances}
M.~K. Mielicki, ``Affordances of graphical and symbolic representations in
  algebraic problem solving,'' Ph.D. dissertation, University of Illinois at
  Chicago, 2015.

\bibitem{herrlinger2017pictures}
S.~Herrlinger, T.~N. H{\"o}ffler, M.~Opfermann, and D.~Leutner, ``When do
  pictures help learning from expository text? multimedia and modality effects
  in primary schools,'' \emph{Research in Science Education}, vol.~47, pp.
  685--704, 2017.

\bibitem{verdi1997organized}
M.~P. Verdi, J.~T. Johnson, W.~A. Stock, R.~W. Kulhavy, and P.~Whitman-Ahern,
  ``Organized spatial displays and texts: Effects of presentation order and
  display type on learning outcomes,'' \emph{The Journal of Experimental
  Education}, vol.~65, no.~4, pp. 303--317, 1997.

\bibitem{mayer2005cambridge}
R.~E. Mayer, \emph{The Cambridge handbook of multimedia learning}.\hskip 1em
  plus 0.5em minus 0.4em\relax Cambridge university press, 2005.

\bibitem{carney2002pictorial}
R.~N. Carney and J.~R. Levin, ``Pictorial illustrations still improve students'
  learning from text,'' \emph{Educational psychology review}, vol.~14, pp.
  5--26, 2002.

\bibitem{fan2015drawing}
J.~E. Fan, ``Drawing to learn: How producing graphical representations enhances
  scientific thinking.'' \emph{Translational Issues in Psychological Science},
  vol.~1, no.~2, p. 170, 2015.

\bibitem{coecke2022kindergarden}
B.~Coecke, D.~Horsman, A.~Kissinger, and Q.~Wang, ``Kindergarden quantum
  mechanics graduates (... or how {I} learned to stop gluing {LEGO} together
  and love the {ZX}-calculus),'' \emph{Theoretical Computer Science}, vol. 897,
  pp. 1--22, 2022.

\bibitem{QxQ}
T.~C. School, ``Qubit by qubit: View past programs,''
  \url{https://www.qubitbyqubit.org/programs}, accessed: 2023-07-30.

\bibitem{OxPhys}
U.~of~Oxford Department~of Physics, ``Quantum club,''
  \url{https://www.physics.ox.ac.uk/engage/schools/secondary-schools/projects-and-mentoring/quantum-club},
  accessed: 2023-07-30.

\bibitem{IQCqsys}
U.~o.~W. Institute~for Quantum~Computing, ``Quantum school for young
  students,'' \url{https://uwaterloo.ca/institute-for-quantum-computing/qsys},
  accessed: 2023-07-30.

\bibitem{economou2020teaching}
S.~E. Economou, T.~Rudolph, and E.~Barnes, ``Teaching quantum information
  science to high-school and early undergraduate students,'' 2020.

\bibitem{seskir2022quantumgames}
Z.~C. Seskir, P.~Migdał, C.~Weidner, A.~Anupam, N.~Case, N.~Davis,
  C.~Decaroli, İlke Ercan, C.~Foti, P.~Gora, K.~Jankiewicz, B.~R.~L. Cour,
  J.~Y. Malo, S.~Maniscalco, A.~Naeemi, L.~Nita, N.~Parvin, F.~Scafirimuto,
  J.~F. Sherson, E.~Surer, J.~R. Wootton, L.~Yeh, O.~Zabello, and M.~Chiofalo,
  ``{Quantum games and interactive tools for quantum technologies outreach and
  education},'' \emph{Optical Engineering}, vol.~61, no.~8, p. 081809, 2022.

\bibitem{lacour2022vqol}
B.~R. La~Cour, M.~Maynard, P.~Shroff, G.~Ko, and E.~Ellis, ``The virtual
  quantum optics laboratory,'' in \emph{2022 IEEE International Conference on
  Quantum Computing and Engineering (QCE)}, 2022, pp. 677--687.

\bibitem{Migdal2022qflytrap}
P.~Migda{\l}, K.~Jankiewicz, P.~Grabarz, C.~Decaroli, and P.~Cochin,
  ``Visualizing quantum mechanics in an interactive simulation {\textendash}
  virtual lab by quantum flytrap,'' \emph{Optical Engineering}, vol.~61,
  no.~08, jun 2022.

\bibitem{chungyuan2022}
Y.-P. Liao, Y.-L. Cheng, Y.-T. Zhang, H.-X. Wu, and R.-C. Lu, ``The interactive
  system of bloch sphere for quantum computing education,'' in \emph{2022 IEEE
  International Conference on Quantum Computing and Engineering (QCE)}, 2022,
  pp. 718--723.

\bibitem{Marshman2022}
E.~Marshman and C.~Singh, \emph{QuILTs: Validated Teaching--Learning Sequences
  for Helping Students Learn Quantum Mechanics}.\hskip 1em plus 0.5em minus
  0.4em\relax Cham: Springer International Publishing, 2022, pp. 15--35.

\bibitem{entanglementball2021}
J.~Marckwordt, A.~Muller, D.~Harlow, D.~Franklin, and R.~H. Landsberg,
  ``{Entanglement Ball: Using Dodgeball to Introduce Quantum Entanglement},''
  \emph{The Physics Teacher}, vol.~59, no.~8, pp. 613--616, 11 2021.

\bibitem{mykhailova2022}
M.~Mykhailova, ``Developing programming assignments for teaching quantum
  computing and quantum programming,'' in \emph{2022 IEEE International
  Conference on Quantum Computing and Engineering (QCE)}, 2022, pp. 688--692.

\bibitem{salehiseskir2022}
O.~Salehi, Z.~Seskir, and I.~Tepe, ``A computer science-oriented approach to
  introduce quantum computing to a new audience,'' \emph{IEEE Transactions on
  Education}, vol.~65, no.~1, pp. 1--8, 2022.

\bibitem{qiskittextbook2021}
J.~R. Wootton, F.~Harkins, N.~T. Bronn, A.~C. Vazquez, A.~Phan, and A.~T.
  Asfaw, ``Teaching quantum computing with an interactive textbook,'' in
  \emph{2021 IEEE International Conference on Quantum Computing and Engineering
  (QCE)}, 2021, pp. 385--391.

\bibitem{khodaeifaal2022}
S.~Khodaeifaal, ``Updated and adapted curriculum and pedagogy of physics with
  the fourth industrial revolution and quantum revolution: From waves
  principles to quantum mechanics fundamentals,'' in \emph{2022 IEEE
  International Conference on Quantum Computing and Engineering (QCE)}, 2022,
  pp. 653--668.

\bibitem{maldonadoromo2022}
A.~Maldonado-Romo and L.~Yeh, ``Quantum computing online workshops and
  hackathon for spanish speakers: A case study,'' in \emph{2022 IEEE
  International Conference on Quantum Computing and Engineering (QCE)}, 2022,
  pp. 709--717.

\bibitem{quantumatlas}
T.~Q. Atlas, ``The quantum atlas,'' \url{https://quantumatlas.umd.edu/},
  accessed: 2023-07-30.

\bibitem{uchicago2022multidisciplinary}
S.~Chitransh, D.~Fischer, N.~Kawalek, N.~LaRacuente, J.~Markman, S.~P. Gaunkar,
  and U.~Zvi, ``A multidisciplinary, artistic approach to broadening the
  accessibility of quantum science,'' in \emph{2022 IEEE International
  Conference on Quantum Computing and Engineering (QCE)}, 2022, pp. 701--708.

\bibitem{hoekzema2007particle}
D.~Hoekzema, E.~van~den Berg, G.~Schooten, and L.~van Dijk, ``The
  particle/wave-in-a-box model in {D}utch secondary schools,'' \emph{Physics
  education}, vol.~42, no.~4, p. 391, 2007.

\bibitem{boe2023secondary}
M.~V. B{\o}e and S.~Viefers, ``Secondary and university students’
  descriptions of quantum uncertainty and the wave nature of quantum
  particles,'' \emph{Science \& Education}, vol.~32, no.~2, pp. 297--326, 2023.

\bibitem{huseby2019observation}
A.~Huseby and B.~Bungum, ``Observation in quantum physics: Challenges for upper
  secondary physics students in discussing electrons as waves,'' \emph{Physics
  Education}, vol.~54, no.~6, p. 065002, 2019.

\bibitem{di2020development}
U.~S. di~Uccio, A.~Colantonio, S.~Galano, I.~Marzoli, F.~Trani, and I.~Testa,
  ``Development of a construct map to describe students’ reasoning about
  introductory quantum mechanics,'' \emph{Physical Review Physics Education
  Research}, vol.~16, no.~1, p. 010144, 2020.

\bibitem{rudolph2017}
T.~Rudolph, \emph{Q Is for Quantum}, 2017.

\bibitem{epiqc2020reversibility}
D.~Franklin, J.~Palmer, W.~Jang, E.~M. Lehman, J.~Marckwordt, R.~H. Landsberg,
  A.~Muller, and D.~Harlow, ``Exploring quantum reversibility with young
  learners,'' in \emph{Proceedings of the 2020 ACM Conference on International
  Computing Education Research}, ser. ICER '20.\hskip 1em plus 0.5em minus
  0.4em\relax New York, NY, USA: Association for Computing Machinery, 2020, p.
  147–157.

\bibitem{greca2003does}
I.~M. Greca and O.~Freire, ``Does an emphasis on the concept of quantum states
  enhance students' understanding of quantum mechanics?'' \emph{Science \&
  Education}, vol.~12, pp. 541--557, 2003.

\bibitem{Penrose}
R.~Penrose, ``Applications of negative dimensional tensors,'' in
  \emph{Combinatorial Mathematics and its Applications}.\hskip 1em plus 0.5em
  minus 0.4em\relax Academic Press, 1971, pp. 221--244.

\bibitem{JS}
A.~Joyal and R.~Street, ``The geometry of tensor calculus {I},'' \emph{Advances
  in Mathematics}, vol.~88, pp. 55--112, 1991.

\bibitem{coecke2011interacting}
B.~Coecke and R.~Duncan, ``Interacting quantum observables: categorical algebra
  and diagrammatics,'' \emph{New Journal of Physics}, vol.~13, no.~4, p.
  043016, 2011.

\bibitem{duncan2020graph}
R.~Duncan, A.~Kissinger, S.~Perdrix, and J.~Van De~Wetering, ``Graph-theoretic
  simplification of quantum circuits with the {ZX}-calculus,'' \emph{Quantum},
  vol.~4, p. 279, 2020.

\bibitem{de2020fast}
N.~de~Beaudrap, X.~Bian, and Q.~Wang, ``Fast and effective techniques for
  {T}-count reduction via spider nest identities,'' in \emph{15th Conference on
  the Theory of Quantum Computation, Communication and Cryptography (TQC
  2020)}.\hskip 1em plus 0.5em minus 0.4em\relax Schloss
  Dagstuhl-Leibniz-Zentrum f{\"u}r Informatik, 2020, arXiv:2004.05164.

\bibitem{de2019techniques}
------, ``Techniques to reduce $\pi$/4-parity-phase circuits, motivated by the
  {ZX} calculus,'' \emph{Electronic Proceedings in Theoretical Computer
  Science}, vol. 318, pp. 131--149, may 2020.

\bibitem{kissinger2019reducing}
A.~Kissinger and J.~van~de Wetering, ``Reducing {T}-count with the
  {ZX}-calculus,'' \emph{arXiv preprint arXiv:1903.10477}, 2019.

\bibitem{huang2023qeczx}
J.~Huang, S.~M. Li, L.~Yeh, A.~Kissinger, M.~Mosca, and M.~Vasmer, ``Graphical
  css code transformation using zx calculus,'' in press 2023.

\bibitem{de2020zx}
N.~de~Beaudrap and D.~Horsman, ``The {ZX} calculus is a language for surface
  code lattice surgery,'' \emph{Quantum}, vol.~4, p. 218, 2020.

\bibitem{kissinger2022phase}
A.~Kissinger, ``Phase-free {ZX} diagrams are {CSS} codes (... or how to
  graphically grok the surface code),'' \emph{arXiv preprint arXiv:2204.14038},
  2022.

\bibitem{khesinGraphicalQuantumCliffordencoder2023}
A.~B. Khesin, J.~Z. Lu, and P.~W. Shor, ``Graphical quantum
  {{Clifford-encoder}} compilers from the {{ZX}} calculus,'' arXiv:2301.02356.

\bibitem{kissinger2022classical}
A.~Kissinger, J.~van~de Wetering, and R.~Vilmart, ``Classical simulation of
  quantum circuits with partial and graphical stabiliser decompositions,''
  \emph{arXiv preprint arXiv:2202.09202}, 2022.

\bibitem{coecke2011phase}
B.~Coecke, B.~Edwards, and R.~W. Spekkens, ``Phase groups and the origin of
  non-locality for qubits,'' \emph{Electronic Notes in Theoretical Computer
  Science}, vol. 270, no.~2, pp. 15--36, 2011.

\bibitem{backens2016complete}
M.~Backens and A.~N. Duman, ``A complete graphical calculus for {S}pekkens’
  toy bit theory,'' \emph{Foundations of Physics}, vol.~46, no.~1, pp. 70--103,
  2016.

\bibitem{gogioso2019dynamics}
S.~Gogioso, ``{A Diagrammatic Approach to Quantum Dynamics},'' in \emph{8th
  Conference on Algebra and Coalgebra in Computer Science (CALCO 2019)}, ser.
  Leibniz International Proceedings in Informatics (LIPIcs), M.~Roggenbach and
  A.~Sokolova, Eds., vol. 139, 2019, pp. 19:1--19:23.

\bibitem{gogioso2017mermin}
S.~Gogioso and W.~Zeng, ``{Generalised {M}ermin-type non-locality arguments},''
  \emph{{Logical Methods in Computer Science}}, vol. {Volume 15, Issue 2},
  2019.

\bibitem{coecke2008interacting}
B.~Coecke and R.~Duncan, ``Interacting quantum observables,'' in
  \emph{Automata, Languages and Programming: 35th International Colloquium,
  ICALP 2008, Reykjavik, Iceland, July 7-11, 2008, Proceedings, Part II
  35}.\hskip 1em plus 0.5em minus 0.4em\relax Springer, 2008, pp. 298--310.

\bibitem{duncan2009graph}
R.~Duncan and S.~Perdrix, ``Graph states and the necessity of {E}uler
  decomposition,'' in \emph{Conference on Computability in Europe}.\hskip 1em
  plus 0.5em minus 0.4em\relax Springer, 2009, pp. 167--177.

\bibitem{kissinger2019universal}
A.~Kissinger and J.~van~de Wetering, ``Universal {MBQC} with generalised
  parity-phase interactions and pauli measurements,'' \emph{Quantum}, vol.~3,
  p. 134, 2019.

\bibitem{QPL-QNLP}
K.~Meichanetzidis, S.~Gogioso, G.~de~Felice, N.~Chiappori, A.~Toumi, and
  B.~Coecke, ``Quantum natural language processing on near-term quantum
  computers,'' \emph{Electronic Proceedings in Theoretical Computer Science},
  vol. 340, pp. 213--229, sep 2021.

\bibitem{lorenz2021qnlp}
R.~Lorenz, A.~Pearson, K.~Meichanetzidis, D.~Kartsaklis, and B.~Coecke,
  ``{QNLP} in practice: Running compositional models of meaning on a quantum
  computer, doi: 10.48550,'' \emph{arXiv preprint arXiv.2102.12846}, 2021.

\bibitem{coecke2020foundations}
B.~Coecke, G.~de~Felice, K.~Meichanetzidis, and A.~Toumi, ``Foundations for
  near-term quantum natural language processing,'' \emph{arXiv preprint
  arXiv:2012.03755}, 2020.

\bibitem{bonchi2014categorical}
F.~Bonchi, P.~Soboci{\'n}ski, and F.~Zanasi, ``A categorical semantics of
  signal flow graphs,'' in \emph{CONCUR 2014--Concurrency Theory: 25th
  International Conference, CONCUR 2014, Rome, Italy, September 2-5, 2014.
  Proceedings 25}.\hskip 1em plus 0.5em minus 0.4em\relax Springer, 2014, pp.
  435--450.

\bibitem{sadrzadeh2013frobenius}
M.~Sadrzadeh, S.~Clark, and B.~Coecke, ``The {F}robenius anatomy of word
  meanings {I}: subject and object relative pronouns,'' \emph{Journal of Logic
  and Computation}, vol.~23, no.~6, pp. 1293--1317, 2013.

\bibitem{wang2023distilling}
V.~Wang-Mascianica, J.~Liu, and B.~Coecke, ``Distilling text into circuits,''
  \emph{arXiv preprint arXiv:2301.10595}, 2023.

\bibitem{lorenz2023causal}
R.~Lorenz and S.~Tull, ``Causal models in string diagrams,'' 2023.

\bibitem{backens2018zh}
M.~Backens and A.~Kissinger, ``{ZH}: A complete graphical calculus for quantum
  computations involving classical non-linearity,'' \emph{arXiv preprint
  arXiv:1805.02175}, 2018.

\bibitem{poor2023completeness}
B.~Poór, Q.~Wang, R.~A. Shaikh, L.~Yeh, R.~Yeung, and B.~Coecke,
  ``Completeness for arbitrary finite dimensions of zxw-calculus, a unifying
  calculus,'' in \emph{2023 38th Annual ACM/IEEE Symposium on Logic in Computer
  Science (LICS)}, 2023, pp. 1--14.

\bibitem{defelicelightmatterZXW}
G.~de~Felice, R.~A.~Shaikh, B.~Po{\'o}r, L.~Yeh, Q.~Wang, and B.~Coecke,
  ``{Light-matter interaction in the ZXW calculus},'' in \emph{Quantum Physics
  and Logic}, in press 2023.

\bibitem{shaikh2022sum}
R.~A. Shaikh, Q.~Wang, and R.~Yeung, ``{How to sum and exponentiate
  Hamiltonians in ZXW calculus},'' in \emph{Quantum Physics and Logic}, in
  press 2022.

\bibitem{coecke2022quantum}
B.~Coecke and S.~Gogioso, \emph{Quantum in Pictures}.\hskip 1em plus 0.5em
  minus 0.4em\relax Quantinuum, September, 2022.

\bibitem{CKbook}
B.~Coecke and A.~Kissinger, \emph{Picturing Quantum Processes. A First Course
  in Quantum Theory and Diagrammatic Reasoning}.\hskip 1em plus 0.5em minus
  0.4em\relax Cambridge University Press, 2017.

\bibitem{hadzihasanovic2018two}
A.~Hadzihasanovic, K.~F. Ng, and Q.~Wang, ``Two complete axiomatisations of
  pure-state qubit quantum computing,'' in \emph{Proceedings of the 33rd Annual
  ACM/IEEE Symposium on Logic in Computer Science}.\hskip 1em plus 0.5em minus
  0.4em\relax ACM, 2018, pp. 502--511.

\bibitem{dowling2003quantum}
J.~P. Dowling and G.~J. Milburn, ``Quantum technology: the second quantum
  revolution,'' \emph{Philosophical Transactions of the Royal Society of
  London. Series A: Mathematical, Physical and Engineering Sciences}, vol. 361,
  no. 1809, pp. 1655--1674, 2003.

\bibitem{coecke2023basic}
B.~Coecke, ``Basic {ZX}-calculus for students and professionals,'' \emph{arXiv
  preprint arXiv:2303.03163}, 2023.

\bibitem{van2020zx}
J.~van~de Wetering, ``{ZX}-calculus for the working quantum computer
  scientist,'' \emph{arXiv preprint arXiv:2012.13966}, 2020.

\end{thebibliography}

\end{document}